\documentclass[english]{article}

\usepackage[utf8]{inputenc}
\usepackage[T1]{fontenc}
\usepackage{babel}
\PassOptionsToPackage{hyphens}{url}\usepackage{hyperref}
\usepackage{amsmath}
\usepackage{graphicx}
\usepackage{tabularx}
\usepackage{pdflscape}
\usepackage{afterpage}
\usepackage{fancyhdr}
\usepackage{rotating}
\usepackage{pgfplots}
    \pgfplotsset{compat=1.14}
    \usepgfplotslibrary{groupplots}
    \usepgfplotslibrary{dateplot}
   \usetikzlibrary{external}
\usepackage{todonotes}
\usepackage{booktabs}
\usepackage{adjustbox}
\usepackage{makecell}
\usepackage{multirow}
\usepackage{float}
\usepackage{listings}
\usepackage{xcolor}
    \definecolor{codegreen}{rgb}{0,0.6,0}
    \definecolor{codegray}{rgb}{0.5,0.5,0.5}
    \definecolor{codepurple}{rgb}{0.58,0,0.82}

\newfloat{lstfloat}{htbp}{lop}
\floatname{lstfloat}{Box}

\lstdefinelanguage{Julia}
  {morekeywords={abstract,begin,break,case,catch,const,continue,do,else,elseif,
      end,export,false,for,function,immutable,import,importall,if,in,
      macro,module,otherwise,quote,return,switch,true,try,type,typealias,
      using,while},
   sensitive=true,
   morecomment=[l]\#,
   morecomment=[n]{\#=}{=\#},
   morestring=[s]{"}{"},
   morestring=[m]{'}{'},
}[keywords,comments,strings]

\lstdefinestyle{mystyle}{
    commentstyle=\color{codegreen},
    keywordstyle=\color{magenta},
    numberstyle=\tiny\color{codegray},
    stringstyle=\color{codepurple},
    basicstyle=\ttfamily\footnotesize,
    breakatwhitespace=false,
    breaklines=true,
    captionpos=b,
    keepspaces=true,
    showspaces=false,
    showstringspaces=false,
    showtabs=false,
    tabsize=2
}

\usepackage[normalem]{ulem}
\usepackage{cancel}

\def\revision#1{\textcolor[rgb]{0.00,0.00,0.00}{#1}}%

\usepackage[binary-units]{siunitx}
\DeclareSIUnit\dBm{dBm}

\usepackage{placeins}

\usepackage{subcaption}

\begin{document}
\tikzexternaldisable

\title{TILES-2018, a longitudinal physiologic and behavioral data set of hospital workers}

\author{
Karel Mundnich\textsuperscript{1{*}},
Brandon M. Booth\textsuperscript{1},
Michelle L'Hommedieu\textsuperscript{1},\\
Tiantian Feng\textsuperscript{1},
Benjamin Girault\textsuperscript{1},
Justin L'Hommedieu\textsuperscript{1},\\
Mackenzie Wildman\textsuperscript{2},
Sophia Skaaden\textsuperscript{3},
Amrutha Nadarajan\textsuperscript{1},\\
Jennifer L. Villatte\textsuperscript{4},
Tiago H. Falk\textsuperscript{5},
Kristina Lerman\textsuperscript{3},\\
Emilio Ferrara\textsuperscript{3},
Shrikanth Narayanan\textsuperscript{1,3}}

\maketitle
\thispagestyle{fancy}

1. Signal Analysis and Interpretation Lab, University of Southern California, Los Angeles, CA, USA,
2. Evidation Health,
3. Information Sciences Institute (USC), Marina del Rey, CA, USA
4. Department of Psychiatry and Behavioral Sciences, University of Washington School of Medicine, Seattle, WA, USA,
5. INRS-EMT, University of Quebec, Montreal, QC, Canada,
{*}corresponding author(s):
Karel Mundnich (mundnich@usc.edu)

\begin{abstract}
We present a novel longitudinal multimodal corpus of physiological and behavioral data collected from direct clinical providers in a hospital workplace. We designed the study to investigate the use of off-the-shelf wearable and environmental sensors to understand individual-specific constructs such as job performance, interpersonal interaction, and well-being of hospital workers over time in their natural day-to-day job settings. We collected behavioral and physiological data from $n = 212$ participants through Internet-of-Things Bluetooth data hubs, wearable sensors (including a wristband, a biometrics-tracking garment, a smartphone, and an audio-feature recorder), together with a battery of surveys to assess personality traits, behavioral states, job performance, and well-being over time. Besides the default use of the data set, we envision several novel research opportunities and potential applications, including multi-modal and multi-task behavioral modeling, authentication through biometrics, and privacy-aware and privacy-preserving machine learning.
\end{abstract} %

\section*{Background \& Summary}

Maintaining a healthy, productive workforce is an increasingly challenging problem in a complex and frenzied world. Optimal job performance relies on worker wellness, and as organizations strive to prepare their workforce for the evolving demands, worker wellness is increasingly important.
Current standards are based on cross-sectional assessment of employee characteristics, often in controlled testing conditions that cannot account for the dynamic nature of working environments and employee performance and are therefore poorly suited for this task \cite{fisher2008if}.
Fortunately, today’s densely instrumented world offers tremendous opportunities for unobtrusive and persistent acquisition and analysis of diverse, information-rich time-series data that provide a multi-modal, spatio-temporal characterization of individuals' actions in, and of, the environment within which they operate. However, the connection between individual and group performance, well-being, and quantitative measurements from sensor data has not been established for such dynamic environments \textit{in the wild}.

To connect job performance-related and well-being-related constructs through self-assessments with data from sensors, we frame well-being and performance within the overarching notion of psychological flexibility.
Psychological flexibility refers to an individual’s capacity to maintain fluid awareness and acceptance of current circumstances and, depending upon available opportunities, take effective action even when experiencing difficult or unwanted thoughts, emotions, and sensations \cite{bond2008influence}.
Psychological flexibility is defined as a primary individual determinant of behavioral effectiveness and well-being \cite{kashdan2010psychological}.
It has been shown that, in the workplace, the degree to which employees are psychologically flexible can have a profound effect on their productivity, well-being, and success \cite{bond2003role, bond2006ability}.
Moreover, the connection between sensor measurements and mental states put forth by the Somatic Marker Hypothesis \cite{damasio1991behavior} suggests that the physiological status of our body (i.e., the somatic marker) is an indispensable part of our cognition and emotion, which are building blocks of our mental states.
The purpose of our research is to connect psychological flexibility, job-performance, and well-being with somatic and bio-behavioral markers using an \textit{in situ} experimental study in a real world workplace.

The TILES-2018 (Tracking IndividuaL performancE with Sensors, year 2018) data set comes from a prospective longitudinal study using intensive multimodal assessment of workers and their environments aimed to understand the dynamic relationships among individual differences, work and wellness behaviors, and the contexts in which they occur.
It aims to support developing and validating sensor-based methods for evaluating worker wellness and job performance over time. To achieve this, we partnered with University of Southern California's Keck Hospital to directly observe 212 workers who volunteered to participate in the study over a 10-week period both at work and outside of work.
Bio-behavioral data were captured continuously and passively throughout the study via wearable devices (including a wristband, a smart undergarment, a clip-on audio-features recorder and Bluetooth-enabled badge, and personal smartphones).
These data streams were matched with environmental and behavioral data streams from Internet-of-Things devices and applications logging personal smartphone usage.
To map sensor data to constructs of interest, participants also completed an initial battery of online surveys and daily surveys designed to assess individual difference variables (e.g., personality, intelligence, socioeconomic status), psychological states and traits (e.g., positive and negative affect, anxiety, stress, fatigue, psychological flexibility,  psychological capital), health and wellness (e.g., sleep, physical activity, cardio exercise, tobacco and alcohol use, health-related quality of life, life satisfaction), and work behaviors (e.g., task performance, organizational citizenship behavior, counterproductive work behavior, work engagement, perceived support and stressors).

This data set provides a unique opportunity for researchers interested in organizational psychology or data sciences in general to perform exploratory and hypothesis-driven investigations regarding the complex, dynamic nature of worker wellness and performance over time.
It is also of interest to signal processing, machine learning, and privacy researchers due to the thousands of hours of sensor data collected across participants in natural real-world \textit{in the wild} settings, that can be used to study and extend current multimodal signal processing methods, perform machine learning inference on psychological states and traits, and study and develop new methods to protect the privacy of users without hindering the richness of such a data set.
Unique strengths of this data set include a rich set of self-assessed psychological constructs coupled with multimodal sensor data, all captured in the wild throughout ten weeks, with high compliance rates of the participants.

The data are available through two records: the Main Data Record, and the Audio Data Record, available at \url{https://tiles-data.isi.edu/}.
\section*{Methods}

\begin{figure}
    \centering
    \includegraphics{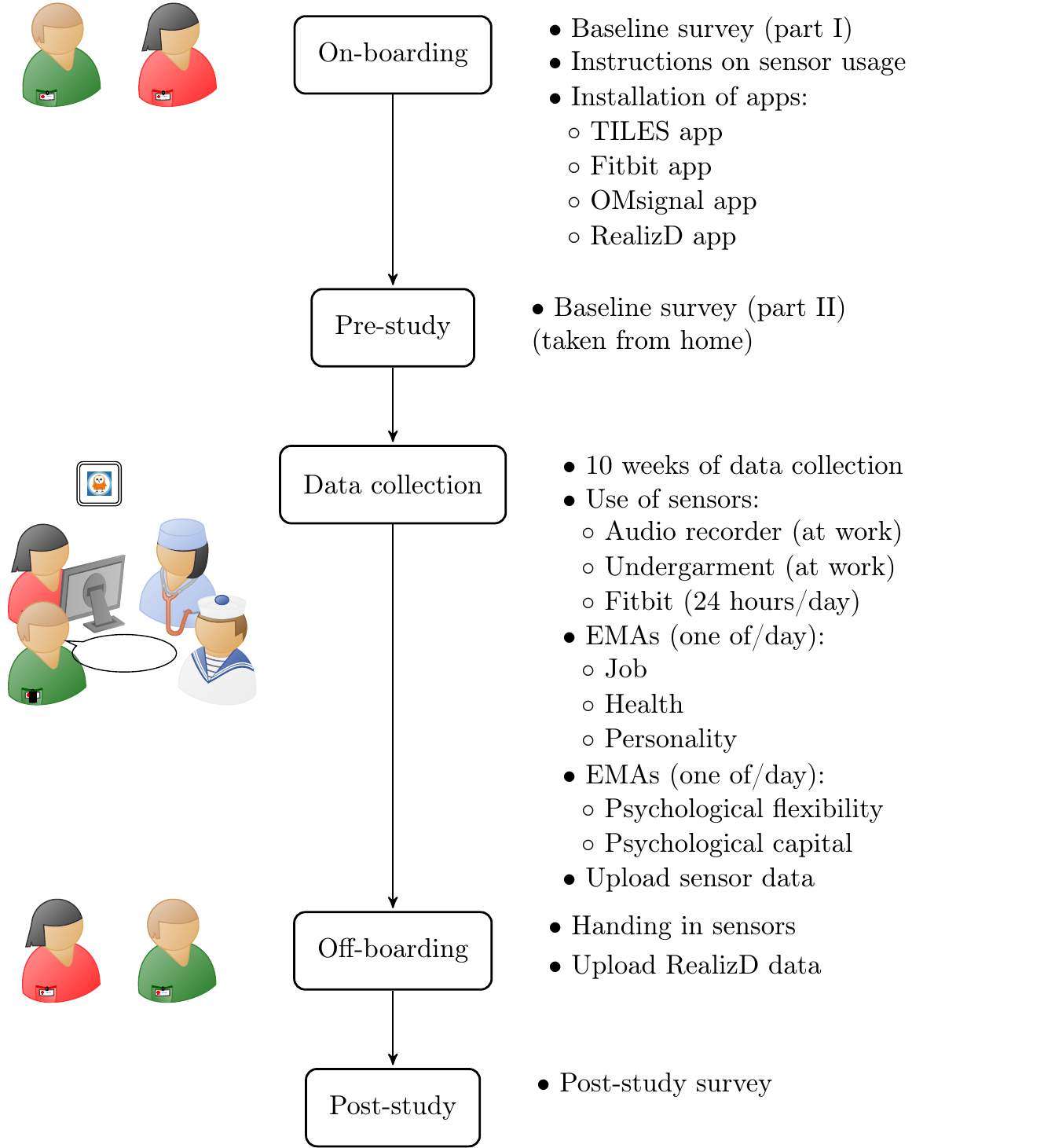}
    \caption{Experimental design. Participants received instructions in a 2-hour on-boarding session, where they completed the first part of the baseline survey and were instructed in the use of sensors and smartphone apps. This session was followed by the second part of the baseline survey and then by 10 weeks of data collection, during which participants wore multiple wearable sensors (wristband, garment and an audio badge) and answered two daily EMAs through their personal smartphones. During the off-boarding session, participants handed in their sensors and finished uploading data and read an audio passage for baseline vocal information. After the sensor data collection, they completed a post-study survey.}
    \label{fig:experimental_design}
\tikzexternaldisable\end{figure}

In this section, we describe the materials and procedures followed to collect the data. An overview of the study is shown in \autoref{fig:experimental_design}.

\subsection*{Location}
The data collection took place at the University of Southern California's (USC) Keck Hospital in Los Angeles, California, in the United States. USC Keck Hospital is an acute care hospital with 401 patient beds throughout 16 nursing units (https://www.keckmedicine.org/about-keck-medicine/keck-hospital-of-usc/). It is located within USC's Health Sciences Campus.

\subsection*{Materials}
This section describes the materials employed in the data collection: Surveys, sensors, and phone applications (apps).

\subsubsection*{Surveys}
To get an understanding of the participants' mental states, traits, and physical and emotional well-being, they were asked to take different surveys throughout the study; these also serve as targets (or labels) for statistical modeling. At the beginning of the study, participants were asked to answer a baseline survey over two different sessions, assessing constructs related to job performance, cognitive abilities, and health. Throughout the data collection, participants answered daily Ecological Momentary Assessments (EMAs) for health, job performance, personality, psychological flexibility, and psychological capital. After the conclusion of the sensor data collection, participants completed a post-study survey. These surveys are described in the following sections, where labels in parenthesis correspond to how the measures are referred to in the data set.

\paragraph*{Baseline Survey}
Due to the length of the baseline survey, it was split into two different sessions. A first part of the baseline survey was administered at the study enrollment session (described in the Study Procedures section) for participants, and assessed demographics and a number of cognitive and psychological constructs pertaining to job performance, cognitive ability, and health. Later, and before the start of the sensor-based data collection, participants answered the second part of the baseline survey at home (or another place of their choice). This survey assessed demographics, health, satisfaction with life, perceived stress, psychological flexibility, work acceptance and action, work engagement, psychological capital, and challenge and hindrance stressors (measures were administered in the above order).

We next describe the scales assessed in the first part of the baseline survey. We give brief descriptions herein (obtained from \cite{MITREwhitepaper}, Table 1) and refer readers to the design and rationale behind this survey in the same document.

\begin{itemize}
\item Demographics (DEMO):
Participants completed a brief demographics survey which assessed sex, age, place of birth, English as the native language, education level, and job-related demographics (e.g., full-time or part-time, industry, tenure in the organization, and income).

\item Cognitive Ability: It was measured using two different scales:
\begin{itemize}
    \item Fluid Intelligence (ABS): Consists of 25 open-ended text entry items, and is scored by adding the sum of correct responses, for a range between 0 and 25.
    \item Crystallized Intelligence (VOCAB): Consists of 40 multiple choice items, with 4 response options each. It is scored by adding the total number of correct answers, for a range between 0 and 40.
\end{itemize}

\item Tobacco Use (GATS):
It was measured using a shortened version of the Global Adult Tobacco Surveys, which consists of 3 items of the form yes/no and quantity questions. The computed scores are tobacco status (\textit{never}, \textit{past}, \textit{current smoker}) and a GATS score, computed by adding tobacco units used in past week (which is $\geq 0$).

\item Alcohol Use (AUDIT):
It was measured using The Alcohol Use Disorders Identification Test \cite{bohn1995alcohol}, which consists of 10 items with yes/no, quantity, and frequency questions. It is scored according to AUDIT instrument scoring guidelines, for a total score in the range 0 to 40.

\item Sleep (PSQI):
It was measured using the Pittsburgh Sleep Quality Index, consisting of 29 items with open-ended response formats as well as structured questions with categorical outcome options. The score is an aggregate sleep quality score, computed according to the PSQI instrument scoring guidelines, for a total score in the range 0-21.

\item Physical Activity (IPAQ):
It was measured using the International Physical Activity Questionnaire, which consists of 27 items of the form yes/no, quantity, and frequency questions. The score is computed using total standardized MET-minutes reported for the prior 7-day period, which is $\geq 0$.

\item Counter-productive Work Behavior (IOD):
It was measured with the Interpersonal and Organization Deviance scale (IOD) \cite{bennett2000development}. It consists of a total of 19 items, separated into two subsets. Each item is a frequency scale ranging from 1 (\textit{never}) to 7 (\textit{daily}).
\begin{itemize}
    \item Interpersonal Deviance (IOD\_ID): Consists of 7 items. The score is computed by adding the responses, for a total score in the range 7 to 49.
    \item Organizational Deviance (IOD\_OD): Consists of 12 items. The score is computed by adding the responses, for a total score in the range 12 to 84.
\end{itemize}

\item Organizational Citizenship Behavior (OCB): Measured using the OCB Checklist (OCB-C) \cite{fox2012deviant}. It consists of 20 items, each being a frequency scale ranging from 1 (\textit{never}) to 5 (\textit{every day}). The score is computed by adding all the responses, for a total score between 20 and 100.

\item Task Performance was assessed using two different measures:
\begin{itemize}
\item In-Role Behavior (IRB) \cite{williams1991job}: Consists of 7 items, each being a Likert scale ranging from 1 (\textit{strongly disagree}) to 7 (\textit{strongly agree}). A scored is obtained by adding all the responses, for a total score between 7 and 49.
\item Individual Task Proficiency (ITP) \cite{griffin2007new}: Consists of 3 items, each being a Likert scale ranging from 1 (\textit{very little}) to 5 (\textit{a great deal}). A scored is obtained by averaging all the responses, for a total score between 1 and 5.
\end{itemize}

\item Personality (BFI-2): It was measured using the Big Five Inventory-2 \cite{soto2017next}. It consists of 60 items, each being a Likert scale ranging from 1 (\textit{disagree strongly}) to 5 (\textit{agree strongly}). Five different scores are computed, all in a range between 1 and 5:
\begin{itemize}
    \item Negative Emotionality (neuroticism): Scored by averaging all the negative emotionality responses.
    \item Conscientiousness: Scored by averaging all the conscientiousness responses.
    \item Extraversion: Scored by averaging all the extraversion responses.
    \item Agreeableness: Scored by averaging all the agreeableness responses.
    \item Open-Mindedness: Scored by averaging all the open-mindedness responses.
\end{itemize}
\item Affect (PANAS):
It was measured using the Positive and Negative Affect Schedule-Expanded Form \cite{watson1999panas}. It consists of 60 items, each being a Likert scale ranging from 1 (\textit{very slightly or not at all}) to 5 (\textit{extremely}). Two different scores were computed, with scores in the range 10 to 50:
\begin{itemize}
    \item Positive Affect (POSAFFECT): Score is obtained by adding the po\-sitive responses.
    \item Negative Affect (NEGAFFECT): Score is obtained by adding the negative responses.
\end{itemize}
\item Anxiety (STAI): It was measured using the State Trait Anxiety Inventory \cite{spielberger1983state}. It consists of 20 items, each being a frequency scale ranging from 1 (\textit{almost never}) to 4 (\textit{almost always}). It is scored by adding sum responses, obtaining a value in the range 20 to 80.
\end{itemize}
The following scales correspond to the second part of the baseline survey, and were assessed on a take-home questionnaire. We include a description of each measurement and a brief rationale.
\begin{itemize}
\item{Demographics (DEMO)}:
Additional demographics assessed several basic characteristics of participants. Specifically, they were asked about race, marital status, pregnancy, number of children living with participants, and housing situation (e.g., rent or own). It also assessed things that were more germane to the particular sample at hand.  This included what position the participant currently held at the hospital from which they were recruited, what specific certifications they have (e.g., nurse practitioner), years in professions, what shift they worked (e.g., day or night), how many hours worked at the organization from which participants were recruited, and amount of over time worked. In addition to this, participants were asked about the length of their commute, mode of transportation used in their commute, do they have another job outside of the one from which they were recruited and if so, how many hours do they work there. Lastly, they were asked if they were currently a student,  their gender, age, place of birth, English as the native language education level, and job-related demographics (e.g., full-time or part-time, industry, tenure in the organization, and income).

\item{Health (RAND)}:
Health was measured using the Rand Health Survey-Short form \cite{hays1993rand}. This assesses eight health domains through 36 self-report items. These domains included
physical function,
role limitations due to physical health,
role limitations due to personal or emotional problems,
general mental health,
social functioning,
bodily pain,
general health perceptions, and
energy/fatigue.
This measure also includes one scale that assesses perceived change in health. Scores are obtained by computing the mean of the items that are associated with each of the domains listed above.

\item{Life Satisfaction (SWLS)}:
The Satisfaction with Life Scale \cite{diener1985satisfaction} is a 5-item measure that aims to assess participants' general satisfaction with life. Participants are to rate the degree to which they agree with each statement on a scale of 1 (\textit{strongly disagree}) to 7 (\textit{strongly agree}). A total score is obtained by taking the average of the 5 items.

\item{Perceived Stress (PSS)}:
The Perceived Stress Scale \cite{cohen1994perceived} is a 10-item scale that aims to assess how often one has experienced stress in the last month. Participants are asked to rate the frequency in which they experience perceive stress on a scale of 0 (\textit{never}) to 4 (very often). After reverse coding the necessary items, a total score is obtained by taking the average of the 10 items.

\item{Psychological Flexibility (MPFI)}:
The Multidimensional Psychological Fle\-xibility Inventory \cite{rolffs2018disentangling} is a 24-item questionnaire that measures both psychological flexibility as well as inflexibility. The 24-item measure is the short form version. 12 items measure flexibility and 12 items measure inflexibility each being assessed on a scale from 1 (\textit{never true}) to 6 (\textit{always true}). The MPFI also measures a number of sub-dimensions:
    \begin{itemize}
        \item Psychological Flexibility (PF): Under flexibility there are sub-dimensions for acceptance, present moment awareness, self as context, defusion, values and committed action.
        \item Psychological Inflexibility (PI): The inflexibility sub-scales include experiential avoidance, lack of contact with the present moment, self as content, fusion, lack of contact with values, and inaction.
    \end{itemize}
Items on this measure ask participants to think about the last two week and to rate the frequency in which they experience the feelings described in each item. PF, PI, and their sub-dimensions are scored by taking the mean of the items that comprised each scale or sub-dimension.

\item{Work Related Acceptance (WAAQ)}: Additionally, psychological flexibility as related to work was measured by the 7-item Work-related Acceptance and Action Questionnaire \cite{bond2013work}.  The WAAQ presents a statement and participants rate the degree to which each statement is true on a scale from 1 (\textit{never true}) to 7 (\textit{always true}). The WAAQ is scored by taking the mean of the items.

\item{Work Engagement (UWES)}:
Work engagement is measured using the Utrecht Work Engagement Scale \cite{seppala2009construct}. Work engagement measure presents 9-items and participants rate the frequency in which they have experienced the feeling described on a scale from 0 (\textit{never}) to 6 (\textit{always}). Then scores are averaged to obtain a total score. There are three sub-scales: vigor, dedication, and absorption.

\item{Psychological Capital (PCQ)}:
It can be thought of as a higher-order construct that is comprised of hope, self-efficacy, resilience, and optimism \cite{youssef2007positive}. It is assessed through the Psychological Capital Questionnaire through a 12-item measure \cite{luthans2007psychological}. The PCQ asks participants the degree to which they agree on a 6-point scale from 1 (\textit{strongly disagree}) to 6 (\textit{strongly agree}).

\item{Challenge and Hindrance Stressors (CHSS)}:
Challenge and Hindrance stressors is measured using a 16-items measure where participants were presented with a statement and asked to rate the degree of agreement or disagreement with the statement \cite{rodell2009can}. 8 items were used to measure challenge stressors and 8 items were used to measure hindrance stressors. Total scores are calculated by computing the mean over all hindrance stressors items and computing separately the mean over all challenge stressor items.
\end{itemize}
\paragraph*{Ecological Momentary Assessments}
The Ecological Momentary Assessments (EMAs) were received twice a day by participants and were divided into two groups.
Note that some scales have a "D" appended to their name compared to the baseline survey to denote its \emph{daily} version.

A first group of EMAs assessed job-related variables, health-related variables, and personality. The job-related questions were asked a total of 31 times during the study (every two days), the health-related questions were asked 35 times during the study (every two days), and the personality-related questions were asked 5 times during the length of the study (every two weeks), with a total of 71 surveys administered over the 10 weeks of the study. Participants received one of these surveys daily. The job, health, and personality surveys were sent either at 6am, noon, or 6pm, and expired 4 hours after they were sent.

Another group of EMAs assessed psychological flexibility and psychological capital. The psychological flexibility form was sent to participants a total of 50 times over the ten weeks (5 times per week), whereas the psychological capital form was received a total of 20 times throughout the same period (2 times per week). Participants received one of these surveys daily. The psychological flexibility and psychological capital EMAs were sent uniformly at random to day shift participants between 11am and 6pm, and between 11pm and 6am for night shift participants. They expired 6 hours after their delivery.

Note that some scales have a "D" appended to their name compared to the baseline survey to denote its \emph{daily} version.

The surveys were implemented using ResearchKit for iOS and ResearchStack for Android (through the TILES app described in Section \nameref{subsubsec:phone-apps}).

The following items were asked daily to participants during ten weeks and were present each at the beginning of each job, health, and personality EMA (base daily survey).

\begin{itemize}
\item{Context measures (CONTEXT)}: These were 4 context questions. The first question asked participants about interactions with other people and the communications channel. The second question asked about the activity in which they were engaged in when they received the survey. The third question asked for current location, and the fourth question asked whether any atypical events had occurred.

\item{Stress (STRESSD)}: Stress was measured daily using a single that read, “Overall, how would you rate your current level of stress?”.

\item{Anxiety (ANXIETY)}: Anxiety was assessed daily using a single which asked, “Please select the response that shows how anxious you feel at the moment”.

\item{Affect (PAND)}: Participants' positive and negative affect were assessed daily using the 10 items from PANAS-Short \cite{mackinnon1999short}.
5 items were used to assess negative affect and 5 items were used to assess positive affect.
\end{itemize}

The purpose of the Job Performance Survey was to assess participants' perceived job performance, and included the following measurements:

\begin{itemize}
\item{Work today (WORK)}: Prior to completing the job performance survey, participants were asked if they had worked 1 or more hours on that day. If participants answered no, they were not shown the job performance survey.

\item{Task performance (ITPD, IRBD)}: Was measured using the same items that were used in the baseline survey described previously.

\item{Organizational citizenship behavior (OCBD)/Counterproductive work behavior (CWBD)}: These were measured using a total of 16 items (DALAL) \cite{dalal2005meta}, with 8 items per scale.
\end{itemize}

The purpose of the Health Survey was to assess a number of health-related variables:

\begin{itemize}
\item{Sleep (SLEEPD)}: Sleep was assessed with a single item that asked participants to specify the number of hours they slept the previous night. Participants were instructed not to confuse this with the number of hours spent in bed.

\item{Physical Activity (EX)}: Physical activity was measured using two questions. Participants were asked to specify the number of minutes of vigorous activity they engaged in yesterday (e.g., sprinting, power lifting). The second, asked participants how many minutes they spent the previous engaging in moderate physical activity (e.g., jogging, biking).

\item{Tobacco Use (TOB)}: Tobacco use was measured using two items. The first asked whether the participant used a tobacco product yesterday and if so, a follow-up question was presented which probed how many times tobacco products were used and what type of product was used.

\item{Alcohol Use (ALC)}: Alcohol use was assessed using 2 items. The first asked whether participants consumed any alcohol yesterday and if they responded yes, they received a question that asked to specify how many beers, wines and spirits they consumed.
\end{itemize}

The purpose of the Personality Survey was to assess the personality:

\begin{itemize}
    \item{Personality (BFID)}: The personality survey uses BFI-10 (shortened version of the BFI-2 used in the baseline survey previously described).
\end{itemize}
 The Psychological Flexibility Survey included context questions and measures of psychological flexibility:

\begin{itemize}
\item{Context Question (Activity)}: The first question asked participants to select from a list the type of activity in which they were engaged in immediately before beginning the survey. Example options included travel or commuting, eating and/or drinking, work, and work-related activities. Participants could also respond ``other'' and specify in text what they were doing.

\item{Context Question (Experience)}:
These items assessed experiences (both pleasant and unpleasant).
The question was provided as a checklist (for positive and negative experiences), such as ``Difficult thoughts of memories'', ``Pleasant physical sensations'', ``Difficult urges or cravings''.

\item{Psychological Flexibility (PF)}: 13 items were included to assess psychological flexibility \cite{bond2008influence}.
Items of the PF survey are divided into 3 sub-scales.
Participants were asked to report how true each statement was about themselves during the last 4 hours. They rated each statement on a scale of 1 (\textit{Never}) to 5 (\textit{Always}). The mean was calculated for all items in each sub-scale for a total score. This scale was created for this study.

\end{itemize}

The Engagement/Psychological Capital Survey  assessed context (base daily survey), engagement, psychological capital, and challenge and hindrance stressors. It is comprised of items that are non-stigmatizing and/or pathologizing, and that have demonstrated large effect sizes on significant outcomes (e.g., employee health and well-being, job performance, job retention and turn-over) \cite{bakker2007job}.

\begin{itemize}
\item{Context questions (Activity)}:
The first question asked participants where they were, and participants selected from a list (e.g., work, home, outdoors, etc.). The second question was the same as the first question participant answered in the context questions for the psychological flexibility questionnaire.

\item{Engagement (Engage)}:
Participants completed a 3-item measure of work engagement \cite{schaufeli2006measurement}. Participants were asked to think about the activity they had just reported doing and how they felt while engaging in that activity. Statements were rated on a scale of 1 (\textit{not at all}) to 7 (\textit{very much}). A mean of the 3 items was computed to create a total score.

\item{Psychological Capital (Psycap)}:
It was measured using 12 items from CPC-12 \cite{lorenz2016measuring}. Participants were instructed to rate each statement based on how much they agreed with it. Items were rated on a scale of 1 (\textit{not at all}) to 7 (\textit{very much}). The mean for all 12 items was used to compute the total score.

\item{Interpersonal Support (IS)}: A subset of 3 items from \cite{tadic2015challenge} are used to assess daily job resources.

\item{Challenge/Hindrance Stressors (CS, HS)}:
A subset of 8 items from the baseline survey measure of Challenge/Hindrance Stressors was used, 4 items to measure each type of stressor \cite{rodell2009can}. Participants were instructed to consider the degree to which they agreed with each statement based on the last day that they had worked, including the day on which they completed the survey. Items were rated on a scale of 1 (\textit{not at all}) to 7 (\textit{very much}).

\end{itemize}

\paragraph*{Post-study Survey}
The Post-study survey is equivalent to the take-home part of the baseline survey, except for not including demographics.

\subsubsection*{Sensing Devices}
The initial goal of the study was to predict self-assessed psychological constructs (obtained through surveys) from sensor data. To this end, we selected a set of wearable and environment-sensing devices to obtain physiological and behavioral information from participants. \autoref{tab:selected_sensors} summarizes the sensors worn by participants and their intended use throughout the study. Details on the sensor selection can be found in \cite{booth2019sensors}.

\paragraph*{Wearable Sensors}
Participants were instructed to wear a Fitbit Charge 2 wristband (\url{https://help.fitbit.com/?p=charge_2}) at all times throughout the duration of the study. Furthermore, at work, they were asked to wear an OMsignal smart garment (\url{https://web.archive.org/web/20181221115159/https://www.omsignal.com/}, a T-shirt for men and a sports bra for women, both discontinued) and a Unihertz Jelly Pro smartphone (\url{https://www.unihertz.com/shop/product/jelly-pro-black-21}, Jelly phone for short) as a lapel microphone (or ``audio badge''). The Jelly phone was programmed to obtain audio features from the raw audio (which was discarded) \cite{feng2018tiles}. In parallel, these Jelly phones also sent Bluetooth packets at \SI{1}{\hertz} over 15\si{\second} windows every minute, to estimate their locations within the building/work place. These packets had a unique 4 bytes identifier for every participant.

\paragraph*{Environmental Sensors}
There were two kinds of environmental sensors: Owl-in-One (\url{https://shop.reelyactive.com/products/owl-in-one-ble}) Bluetooth data hubs and Minew sensors (\url{https://en.minewtech.com/sensor.html}). The Owl-in-Ones were used to estimate participant proximity \revision{to these} by capturing the signal strength of Bluetooth packets from the Jelly phones that participants wore in the hospital and to collect environmental data sent over Bluetooth by Minew sensors.

The Owl-in-Ones were installed in fourteen nursing units (spread over seven of the building floors) and two hospital labs.
A total of 244 Owl-in-Ones were installed, about \SIrange[range-units = repeat]{1.5}{2.0}{\metre} above the floor depending on space availability on wall areas near power outlets. Each nursing unit was equipped with an Owl-in-One sensor in these four room types: patient room, nursing station, lounge, and medication room. These different rooms were selected after observing the behavioral patterns of nurses during their shifts (by talking to nursing directors of Keck Hospital and shadowing nurses throughout a workday). Each Owl-in-One was labelled with the study logo, and the phrase ``This is a data hub for the TILES study. For more information, please visit \url{https://sail.usc.edu/tiles}''.

\begin{figure}
    \centering
    \includegraphics[width=0.6\textwidth]{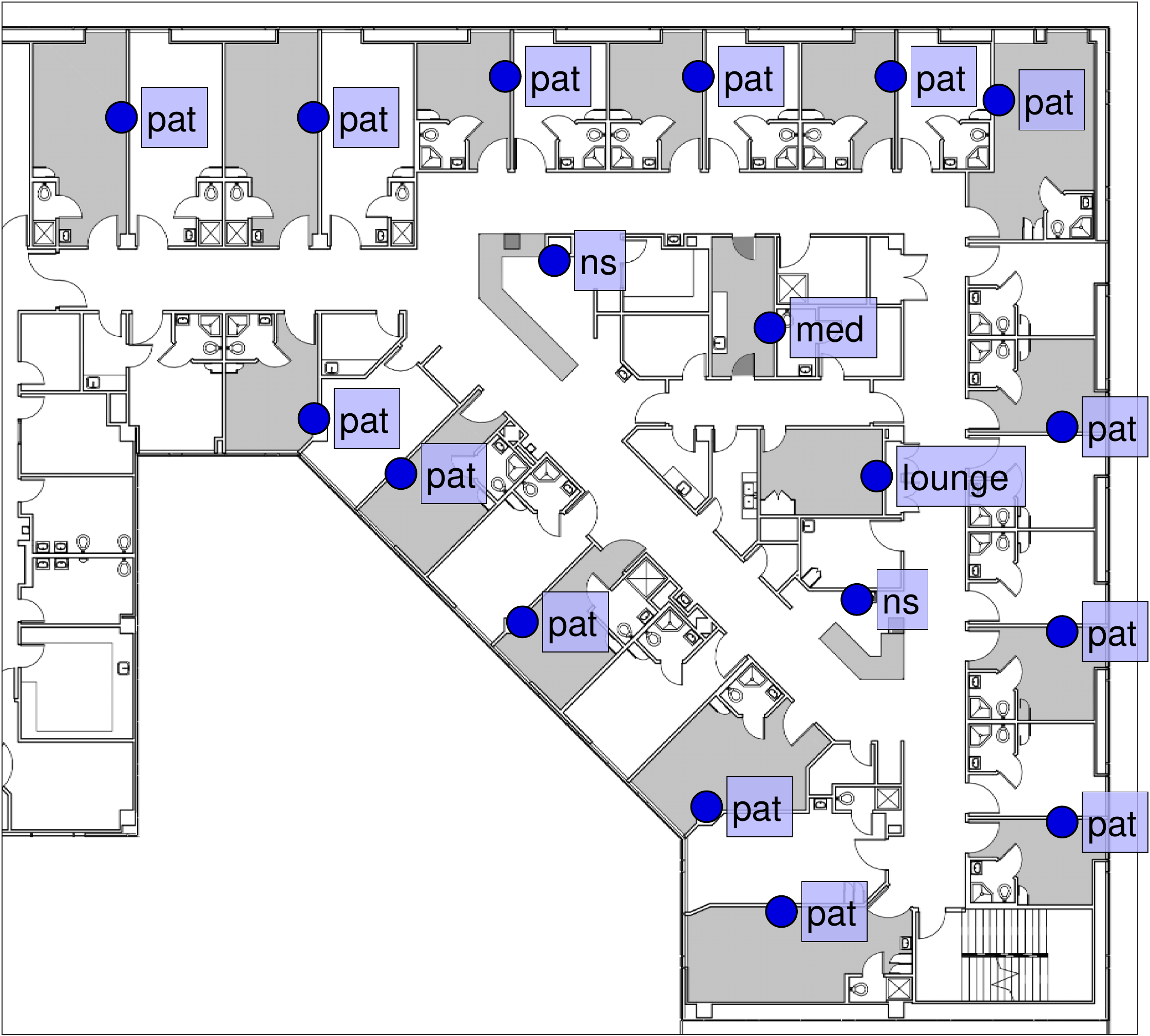}
    \caption{Positioning of Owl-in-Ones (Bluetooth hubs) within a nursing unit. Rooms shaded in gray contain an Owl-in-One, while the blue circles show their exact locations. \texttt{pat} denotes a patient room, \texttt{ns} a nursing station, \texttt{med} a medication room, and \texttt{lounge} represents the lounge or break area for the workers.}
    \label{fig:owl-locations}
\tikzexternaldisable\end{figure}

One Owl-in-One was installed in every other patient room, one in every medication room, one in every lounge, and between one and four in nursing stations, depending on the size, layout, and availability of power outlets. In the hospital labs, one Owl-in-One was installed in every lounge, and at least one in each major room (e.g., blood lab, micro-bio lab, shipping/receiving, patient lobby, etc.) depending on the room size and power outlet availability. \autoref{fig:owl-locations} shows an example of Owl-in-One placements in a nursing unit.

Through information collected from Minew sensors, the Owl-in-Ones also captured (door) motion information, humidity, temperature, and light information across the hospital. Two light (E6) and temperature/humidity (S1) Minew sensors were installed in each nursing unit and each laboratory.  These sensors were placed in open areas near the main hallways and within one foot of an Owl-in-One sensor. In the nursing units, one pair of E6/S1 sensors was installed in the nursing stations nearest and farthest from the unit entrance. In the labs, one pair was located near the lab entrance and the other in a frequently occupied open room away from the entrance. Minew motion sensors (E8) were placed on the top outer corner of doors and captured information pertaining to foot traffic through the doorway.  One motion sensor was placed on each medicine room door in the nursing units.  No sensor was placed on the lounge room doors because they remained open at all times, and none were placed on the unit entrance/exit doors due to fire safety restrictions.  In the labs, one motion sensor was placed on the main entrance door and one on the lounge door. A total of 52 motion sensors, 63 light sensors, and 63 temperature/humidity sensors were installed throughout the hospital.

\begin{table}[tb]
    \centering
    \begin{tabular}{@{ }lll@{ }}
    \toprule
        \textbf{Sensor} & \textbf{Measurements} & \makecell{\textbf{Instructed use/}\\\textbf{Sensing times}} \\
    \midrule
        Fitbit Charge 2 & \makecell[l]{PPG-based heart rate,\\ step count, sleep} & 24 hours/day\\
    \midrule
        OMsignal garments & ECG, breathing, motion & At work (\SI{12}{hour} shifts)\\
    \midrule
        Unihertz Jelly Pro & \makecell[l]{Audio features} & At work (\SI{12}{hour} shifts) \\
    \midrule
        RealizD & Personal phone usage & 24 hours/day\\
    \midrule
        TILES app & \makecell[l]{Ecological Momentary\\ Assessments (EMAs)} & \makecell[l]{Upon request (push\\ event)}\\
    \midrule
    	Web browser & \makecell[l]{Pre and post-study\\surveys} & \makecell[l]{Twice during the\\ study}\\
    \midrule
        reelyActive Owl-in-One & \makecell[l]{Received signal strength\\ of Bluetooth packets} & \makecell[l]{Installed at USC’s Keck\\ Hospital, 24 hours/day}\\
    \midrule
        Minew E6 & Light & \makecell[l]{Installed at USC’s Keck\\ Hospital, 24 hours/day}\\
    \midrule
        Minew E8 & \makecell[l]{Motion (accelerometer\\ \& gyroscope)} & \makecell[l]{Installed at USC’s Keck\\ Hospital, 24 hours/day}\\
    \midrule
        Minew S1 & Temperature, humidity & \makecell[l]{Installed at USC’s Keck\\ Hospital, 24 hours/day}\\
    \bottomrule
    \end{tabular}
    \caption{Selected sensors with a summary of measurements (output) and instructed use or sensing times. The first three sensing streams were obtained directly from participants through wearable sensors and apps installed in their personal smartphones. All surveys were obtained by direct input of participants on their personal smartphones or a web browser. The last four sensing streams were obtained by placing sensors in the hospital. PPG: photoplethysmography, ECG: electrocardiography.
    }
    \label{tab:selected_sensors}
\end{table}

\subsubsection*{Phone apps}\label{subsubsec:phone-apps}
Several phone apps were installed, with informed consent, on the participants' personal smartphones, for interaction with sensors, data uploading, to receive surveys, and to communicate with the research team.

\paragraph*{TILES app}{
This app was custom-developed for the TILES study and was used both for data collection and for communication with participants throughout the enrollment and data collection periods. It is available for both Android and iOS (see Section \nameref{subsec:code-availability} for details). The EMAs were administered via the TILES app. Participants received a push notification when the EMAs were delivered and again thirty minutes before it expired if it had not yet been completed. Bi-directional communication was enabled via the TILES app as well. Participants could contact the research team at any time through the \textit{Contact Info} tab. The app also contained a \textit{Frequently Asked Questions} (FAQs) page which was updated in real time during the study as common questions were identified. In return, participants were notified via push notifications, and the via activity feed within the app of any non-compliance and were reminded to sync each device with its companion app.
}

\paragraph*{Fitbit app}{
The Fitbit app is a third party app that was used to pair the Fitbit wristband with each participant's personal smartphone using Bluetooth. Participants could visualize the data collected through their Fitbit wristband in this app, and could sync their data with Fitbit's servers.
}

\paragraph*{OMsignal app}{
The OMsignal app is a third party app that was used to start and stop the recording of the OMsignal garments, update the firmware of OMsignal garments if necessary, and sync the data to OMsignal's servers.
}

\paragraph*{RealizD app}{
RealizD is a third party smartphone application (no longer developed) for iOS and Android that records screen-on time and phone pickups. Data reported by RealizD takes the form of a timestamp marking the start screen-on session and the duration of that session in seconds.
}

\subsection*{Study Procedures}
In this section we describe the mechanisms through which participants were deemed eligible and later recruited and enrolled in the study. We also describe the data collection process. All these steps were conducted in accordance with USC’s Health Sciences Campus Institutional Review Board (IRB) approval \revision{(study ID HS-17-00876)}. We present an overview of the study in \autoref{fig:experimental_design}.

\subsubsection*{Requirements for eligibility}
All volunteer participants were recruited from the University of Southern California's (USC) Keck Hospital. To participate, subjects were required to (a) be employed by the hospital and work, on average, at least 20 hours a week, (b) have exclusive access to an internet and Bluetooth-enabled mobile phone running Android 4.3 or higher or iOS 8 or higher for the 10 weeks of participation, (c) have exclusive access to a personal e-mail for the 10 weeks of participation, (d) have access to WiFi at home for the duration of the 10 week study, (e) be proficient in both speaking and reading English, and (f) be capable of wearing wearable sensors in a way that allows data to be collected and transmitted to the research team.

\subsubsection*{Recruitment}
Participants were recruited using multiple methods, including (a) e-mails to employees from leaders within Keck Hospital informing them about the study and how to sign up, (b) attending employee meetings to inform employees about the study, (c) posting flyers in different parts of the hospital where employees would be likely to see them, (d) information tables set up in the cafeteria, where potential participants could learn more about the study and sign up. Participants who had indicated interest but had not completed the sign-up process were texted by one of the principal investigators to support completion of the sign-up process.

After completing a screening questionnaire to check eligibility, potential participants were sent a text message with a link to download the TILES app. The TILES app then walked them through identity verification, informed consent, downloading and syncing the necessary additional apps, and finally signing up for an in-person enrollment session.

Through the above methods, 365 individuals indicated interest in participating by completing a brief screening questionnaire and were found to be eligible. Of these 365 individuals, 212 participants provided their consent to participate in the study, \revision{while 153 did not complete the on-boarding procedures}. Participants were recruited in three waves, each with different start and end dates. \autoref{tab:waves} summarizes the dates and number of participants per wave. Over the course of the study, eight participants chose to drop out, due to various reasons, such as a sensor becoming uncomfortable or no longer wanting to receive daily surveys. The data of these participants has been kept in the dataset.

\subsubsection*{Participant enrollment session}
After providing their consent to participate, interested individuals signed up for a two-hour in-person enrollment session at the hospital through the TILES app. Upon arrival at the enrollment session, each participant was assigned a unique participant ID. During the first hour, participants completed part I of the baseline survey, under the supervision of a trained research team member. During the second hour, participants received their package of wearable sensors and instructions for use. Each participant received three wearable sensors along with a USB charging hub and two micro USB cables for charging, to help participants streamline the process of charging the sensors. The TILES app sent participants links to download all the necessary apps: Fitbit, OMsignal, and RealizD.

Participants were instructed to wear three sensors (a Fitbit Charge 2, an OMsignal garment, and a Unihertz Jelly Pro smartphone) that collected physiological and behavioral data over a 10-week period. We describe the instructions given to participants in the following paragraphs. \autoref{tab:selected_sensors} shows a list of the sensing streams and their instructed use. \revision{In addition, participants were instructed to fill part II of the baseline survey at home.}

\paragraph*{Daily Surveys}{Participants were informed from the first day of data collection they would start receiving one text message each day they were enrolled in the study. The text message contained a link to the job, health, or personality EMAs that they were expected to complete that day. Participants were instructed to complete the survey as soon as safely possible once they received the text message. A second daily EMA with psychological flexibility or capital surveys was received via a push notification on the participant's phone and contained similar instructions.

The EMAs took no more than 15 minutes to complete, and on most days the survey could be completed in around 5 minutes. Participants who worked on the night shift received the first EMA (job, health, or personality) at either 6pm, 12am, or 6am and participants who worked the day shift received the job, health, or personality EMAs at either 6am, 12pm, or 6pm. Participants were informed that they had 6 hours to complete each survey and they would receive a reminder notification from the TILES app 30 minutes before the link expires if the survey was not complete. The research team then distributed a calendar of the 10-week data collection period with a schedule of when to expect the daily survey each day. For the second (psychological flexibility or capital) EMAs, night shift participants received the surveys at a random time between 11PM and 6AM and were given 6 hours to complete the survey once it had been sent. Day shift participants received these surveys at a random time between 11AM and 6PM and were given 6 hours to complete the survey once it had been sent. All participants would receive a reminder via a push notification 30 minutes before the survey closed to remind them to complete the survey.}

\paragraph*{Fitbit Charge 2}{The first wearable sensor distributed to participants was the Fitbit Charge 2. Participants were asked to wear this sensor on their non-dominant hand day and night throughout participation in the study. To properly set-up this sensor, each participant created a Fitbit account and registered the Fitbit Charge 2 as a new device as well as synced the Fitbit app to the TILES app. When prompted by the Fitbit app, participants were asked to give Bluetooth permissions and deny location permissions.
}

\paragraph*{OMsignal garments}{Next, participants were given an OMbox and OMsignal garments; men were given five shirts and women were given three bras. The OMbox contains the hardware and software to process, collect, and transmit the information. Participants were asked to charge the sensor prior to each work shift, then connect it to the OMsignal garment, wear the OMsignal garment with the OMbox attached during their work shifts at the hospital, and start OMsignal recordings in the OMsignal app installed in their phones at the beginning of each work shift and stop, save and upload the recording at the end of each work shift. During the enrollment session, each participant paired his/her OMsignal box to his/her account (created through the TILES app) on the OMsignal app on their mobile phone, practiced connecting the OMsignal box to the garment, and saving an OMsignal recording. At the beginning of the data collection, there was no version of the OMsignal app for Android. As a solution, we provided an iPod Touch to each participant with an Android personal smartphone with the OMsignal app installed. This way, they could start and stop recordings and upload the data using WiFi. The research team also helped set up location-based reminders on the iOS devices to help participants remember to start and stop OMsignal recordings when arriving at the hospital as well as leaving.}

\paragraph*{Unihertz Jelly Pro}{Participants were given an Unihertz Jelly Pro phone (running Android 7.0). These were either clipped to participants' clothing near the neckline or placed in a shirt pocket. The cases of the Jelly phones were modified, such that the microphone pointed upwards, as described in \cite{feng2018tiles}, to better capture the speech data from the wearer. Participants were asked to charge the Jelly phone prior to each work shift, unlock the Jelly phone, check that the TILES Audio app \cite{feng2018tiles} was running, and upload the audio data at the end of each work shift by pressing the \textit{UPLOAD DATA} button in the TILES Audio app.  Each Jelly phone was linked to the TILES app on each participant’s mobile device by scanning a QR code in the TILES app. When prompted by the Jelly phone TILES Audio app, participants were asked to enable permissions (e.g., allowing TILES Audio app to run in the background, access to photos even though the camera was not used, but access was needed for the proper functioning of the app, etc.) and disable location-related services. Additionally, participants were informed of the Jelly Phone TILES Audio app’s disable feature (to stop recording audio features) and instructed on how to use this function.}

\paragraph*{RealizD app}{Lastly, participants downloaded the RealizD app on their smartphone and were informed that this app would track how often the phone was picked up and for how long. Participants did not need to interact with the RealizD app during their participation, since it ran in the background.}

\paragraph*{Phone permissions}{
For the RealizD app to work, participants were asked to allow location permissions. Participants were also asked to keep WiFi and Bluetooth turned on on their personal mobile phones throughout their participation in the 10-week data collection period.}

\paragraph*{Environmental sensors}{Finally, participants learned about the environmental sensors that were placed around the hospital and informed that no participant interaction with these sensors was required.}

\subsubsection*{Completing the Pre-Study Survey}
Following completion of their enrollment session, participants were emailed a link to complete this survey, administered on the online survey platform REDCap.

\subsubsection*{Data Collection}
The 10-week data collection took place in three different participant waves. The data collection periods and number of participants per wave are shown in \autoref{tab:waves}.

\begin{table}[tb]
    \centering
    \begin{tabular}{crrrcc}
        \toprule
        \textbf{Wave} & \textbf{Start date} & \textbf{End date} & $\mathbf{n}$ & \textbf{\% of total} & \textbf{Dropouts} \\
        \midrule
        1 & March 5, 2018 &  May 14, 2018 &  52 & 24.5 & 2\\
        2 & April 9, 2018 & June 18, 2018 & 116 & 54.7 & 6\\
        3 &   May 4, 2018 & July 14, 2018 &  44 & 20.8 & 0\\
        \midrule
        \textbf{Total} &&               & 212 && 8\\
        \bottomrule
    \end{tabular}
    \caption{Data collection implementation. This table shows the start and end dates of each wave, with corresponding number of participants at the beginning of each wave and dropouts. Specific dropout dates are given in the folder {\tt metadata/participant-info} (see Section Main Data Record).}
    \label{tab:waves}
\end{table}

\subsubsection*{Off-boarding Session}
After the 10-week data collection from sensors and daily surveys ended, participants attended an in-person off-boarding session, which typically lasted between 15 to 20 minutes.
During this session, participants exported mobile application data to members of the research team and returned their wearable sensors (except for Fitbit, see Section Incentives).

\subsubsection*{Completing the Post-Study Survey}
Following completion of their off-boarding session, participants were emailed a link to complete a survey administered on the online survey platform REDCap. This survey was identical to part II of the baseline survey; the only difference is that the demographics survey was removed and a study feedback survey was administered. This survey took approximately 30 minutes to complete. This concluded participant study procedures.

\subsection*{Incentives Structure}
\begin{table}[tb]
    \centering
    \begin{tabular}{lr}
    \toprule
        \textbf{Stage} & \textbf{Incentive} \\
    \midrule
        Enrollment session        & \$75 \\
        Baseline survey (part II) & \$75 \\
        Data collection           & (up to, weekly) \$25 \\
        Post-study survey         & \$75 \\
    \bottomrule
    \end{tabular}
    \caption{Summary of monetary incentives. Participants were paid after the completion of different stages throughout the study.}
    \label{tab:incentives_per_stage}
\end{table}

A novel incentive scheme was developed for the TILES study to encourage compliance. Study participants were awarded with monetary incentives (\autoref{tab:incentives_per_stage}) and points for study-related activities, proportionate to the time required to complete each activity. These points later translated to monetary awards. The number of points awarded for each activity is summarized in \autoref{tab:weekly_points}. A survey was considered completed if the participant went through all the survey (but they could skip questions). Note that for at least three consecutive days of Fitbit data, the participant received a $2\times$ boost on points received for wearing the Fitbit. Points were converted to monetary compensation on a weekly cadence, according to a set of thresholds noted in \autoref{tab:weekly_reward_cuttoffs}. The expected use of OMsignal garments and Jelly phones was 3 days a week for most of the participant population, so points for wearing and syncing these devices were added to the incentives schemes as bonuses.

In addition to weekly gift cards as incentive, points were accumulated throughout the duration of the study and grand prizes were awarded to the top three point earners per wave.  Each participant's current point total and ranking were displayed in the TILES app activity feed. Bonus points were awarded for various activities, as summarized in \autoref{tab:bonus points}. The first, second, and third place point earners across each wave were awarded \$250, \$200, and \$100, respectively.

Participants that finished the 10-week data collection period also kept the Fitbit Charge 2 that they wore during the study.

\begin{table}[tb]
    \centering
    \begin{tabular}{llr}
    \toprule
        \textbf{Kind} & \textbf{Action} & \textbf{Points} \\
    \midrule
    \multirow{4}{*}{Base} & Open the TILES app                                     &  5 \\
                          & Complete EMA (group I -- job, health, personality)     & 10 \\
                          & Complete EMA (group II -- psy. flexibility \& capital) & 10 \\
                          & Wear and sync Fitbit                                   & 10 \\
    \midrule
    \multirow{5}{*}{Bonuses} & Multiplier for 3 consecutive days of Fitbit data    & $\times \phantom{1}2$ \\
                             & Reach at least 275 weekly points                    & 20 \\
                             & Earn more points than the previous week             & 20 \\
                             & Wear and sync OMsignal device at least 2 days       & 20 \\
                             & Wear and sync Jelly phone at least 2 days           & 20 \\
    \bottomrule
    \end{tabular}
    \caption{Weekly points given to participants. The points were assigned based on the completion of the tasks.}
    \label{tab:weekly_points}
\end{table}{}

\begin{table}[tb]
    \centering
    \begin{tabular}{cc}
    \toprule
       \textbf{Minimum \# points} & \textbf{Gift card amount} \\
    \midrule
        250 & \$25 \\
        200 & \$20 \\
        150 & \$15 \\
        100 & \$10 \\
    \bottomrule
    \end{tabular}
    \caption{Weekly reward cutoffs. Weekly points awarded for compliance in sensor usage and answering of the surveys were translated into monetary rewards.}
    \label{tab:weekly_reward_cuttoffs}
\end{table}

\begin{table}[tb]
    \centering
    \begin{tabular}{llc}
    \toprule
        \textbf{Instance} & \textbf{Action} & \textbf{Points} \\
    \midrule
        \multirow{2}{*}{On-boarding}    & Download and install the TILES app                & 50 \\
                                        & Authorize Fitbit access                           & 50 \\
    \midrule
       \multirow{4}{*}{Weekly}          & Reach at least 275 weekly points      & 20 \\
                                        & Earn more points then the previous week           & 20 \\
                                        & Wear and sync OMsignal device at least two days   & 20 \\
                                        & Wear and sync Jelly Pro device at least two days  & 20 \\
    \midrule
        Off-boarding                    & Export RealizD data                               & 50 \\
    \bottomrule
    \end{tabular}
    \caption{Bonus points scheme for study participation stages and milestones. Participants received weekly points by wearing the sensors and answering the surveys. These were converted to weekly monetary rewards (\autoref{tab:weekly_reward_cuttoffs}) and added to a global ranking that awarded prices by the end of the study.}
    \label{tab:bonus points}
\end{table}

\subsection*{Data acquisition and flow}
\begin{figure}[t]
    \centering
    \makebox[\textwidth][c]{\resizebox{1.25\textwidth}{!}{
        \includegraphics{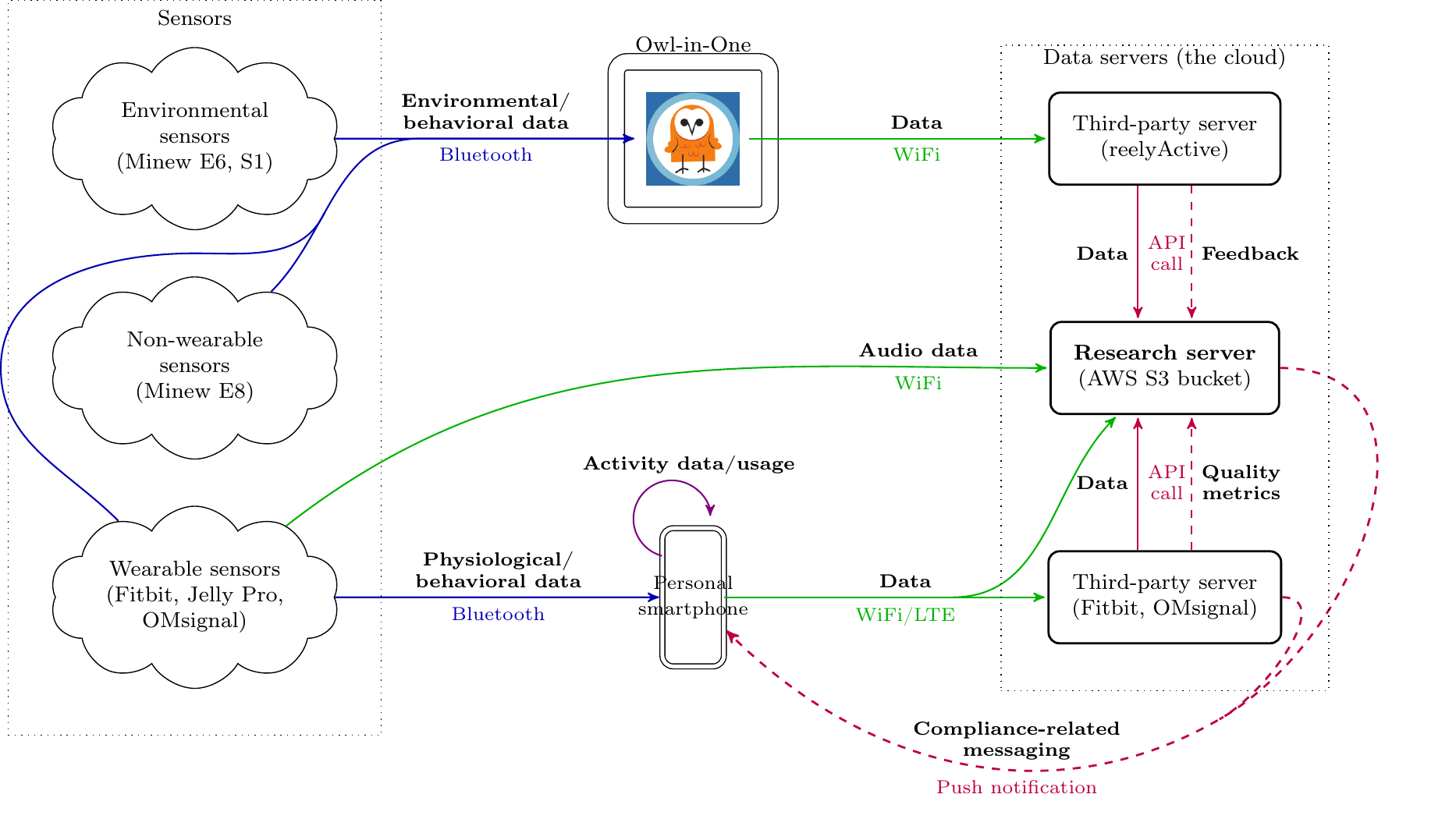}
    }}
    \caption{Data flow. This diagram shows the data flow from the sensors given to participants, the sensors placed at USC's Keck Hospital, and smartphones to the research server, where the data is stored for long-term use.}
    \label{fig:data_flow}
\tikzexternaldisable\end{figure}

\autoref{fig:data_flow} depicts the architecture for the data collection from sensors. On the left column, we have all possible wearable and environmental sensors. Wearable sensors such as Fitbit and OMsignal garments connect to the participants' personal smartphones using Bluetooth. The data are uploaded to a third-party server using an available wireless internet connection (WiFi or LTE).

The Jelly phones (used here as audio-features recorders) uploaded data to the research server directly using WiFi \textit{only}. The Jelly Pro smartphones given to participants also sent Bluetooth packets programmed with a unique identifier that were captured by the Owl-in-One hubs installed throughout the hospital. These packets were combined with their received signal strength indicator (RSSI) computed by the Owl-in-Ones.

The Owl-in-Ones also received data from environmental sensors. Data from both Minew sensors and Jelly phones were sent through Keck Hospital's public WiFi network to reelyActive's servers over UDP, from which the data were collected using the Pareto API \cite{ParetoAPI} over HTTPS. These data were stored securely in the research server after filtering the data to contain only Bluetooth packets generated by our sensors.

Data were also collected directly through the participants' personal smartphones through the TILES app and the RealizD app. The TILES app uploaded data directly to the research server while the RealizD app uploaded data to the RealizD server and that data was later pulled to the research server. The research server (code available at \url{https://github.com/usc-sail/tiles-data-collection-pipeline/}) consists of a RESTful API hosing a series of endpoints to collect push-type data streams (e.g. Owl-in-One, TILES app) in addition to a suite of tasks to fetch pull-type data streams (e.g. Fitbit, OMsignal).

\subsection*{Data Preprocessing} \label{sec:data_preprocessing}
\subsubsection*{Survey data}
Once the data collection period ended, the baseline survey and EMAs were scored using R scripts (available at \url{https://git.io/JePgE}).

Data for the baseline survey were stored in a table where each column represents a single survey question \revision{or metadata variable} and each row represents a single participant.
In contrast, data for the various EMAs were measured daily and stored in multiple files, where each row contains the answers of a single participant to a survey.
The files are split by shift (day/night), date, survey kind, and time it was administered.
\revision{
Surveys left unanswered by participants were added as empty surveys later on.
All these files were aggregated and curated to obtain three files, one for the first group of EMAs (job, health, and personality in addition to base), one for the psychological flexibility group and one for the psychological capital group (curation scripts are available in the folder \texttt{src/curation/} of the companion code).
We have removed most of the raw questions from the EMAs to preserve participants' privacy (except for those included in \autoref{tab:data_preprocessing:surveys:ema_anonymization}), but have kept the aggregated scores.
We also list in \autoref{tab:data_preprocessing:surveys:demo_anonymization} the demographic variables that underwent additional curation to prevent de-identification.
}

\revision{
Free text responses in EMAs have been manually annotated into categories.
Three questions are concerned: location when answering, activity engaged in right before answering, and atypical events that happened or are expected to happen.
Since some of these categories are subjective, we have between 2 and 5 annotations (one per annotator) for each text response.
Each text response can have between 1 and 3 categories associated with by annotators.
Fusion of annotation is then performed and the top 3 categories appearing at least twice are reported alongside the frequency of the category in the annotations (e.g. if 2 out of 5 annotators use a category, that category has frequency $2/5=0.4$).
We refer the reader to the \texttt{README} file in the dataset for further details on those categories and how they are reported in the data.
}
\begin{table}[htb]
    \centering
    \revision{
    \begin{tabular}{lll}
        \toprule
        \textbf{Variable} & \textbf{Operation} & \textbf{Details} \\
        \midrule
        \multicolumn{3}{c}{\textbf{EMAs: job \& health \& personality base survey}} \\
        \midrule
        \texttt{context2\_TEXT} & categorized & see README \\ 
        \texttt{context3\_TEXT} & categorized & see README \\ 
        \texttt{context4\_TEXT} & categorized & see README \\ 
        \midrule
        \multicolumn{3}{c}{\textbf{EMAs: psychological flexibility context survey}} \\
        \midrule
        \texttt{Activity} & categorized & same as \texttt{context2\_TEXT} \\
        \midrule
        \multicolumn{3}{c}{\textbf{EMAs: psychological capital context survey}} \\
        \midrule
        \texttt{Location} & categorized & same as \texttt{context3\_TEXT}  \\
        \texttt{Activity} & categorized & same as \texttt{context2\_TEXT}  \\
        \bottomrule
    \end{tabular}
    \caption{EMA surveys anonymization. We have not included in this table the variables whose values did not require changes.}
    \label{tab:data_preprocessing:surveys:ema_anonymization}
    }
\end{table}

\begin{table}[htb]
    \centering
    \revision{
    \centerline{
    \begin{tabularx}{0.75\paperwidth}{lllll}
        \toprule
        \textbf{Variable} & \textbf{Operation} & \textbf{Details} \\
        \midrule
        \multicolumn{3}{c}{\textbf{Baseline Part I}} \\
        \midrule
        \texttt{age} & binned & $<30$, $\geq 50$, 5 years bins in between \\
        \texttt{country} & removed & \\
        \texttt{englyrs} & binned & $[0,25)$, $[25,35)$, $\geq 35$ years \\
        \texttt{educ} & coarsened & $\leq$ college merged, college degree kept, $\geq$ graduate school merged \\
        \texttt{jobstat} & removed & \\
        \texttt{occup} & removed & \multirow{2}{*}{replaced by part II baseline (\texttt{currentposition})} \\
        \texttt{occup\_TEXT} & removed & \\
        \texttt{quantsup} & coarsened & $<5$, $\geq 5$ people \\
        \texttt{size} & removed & misunderstood question \\
        \texttt{duration} & coarsened & 5, 6, 7 and 8 years merged into $[5,8]$ \\
        \texttt{income} & coarsened & $[\$0,\$25\text{k})$ and $[\$25\text{k},\$50\text{k})$ merged \\
        \midrule
        \multicolumn{3}{c}{\textbf{Baseline Part II}} \\
        \midrule
        \texttt{race} & coarsened & white/caucasian, asian, n.a., other \\
        \texttt{relationship} & coarsened & divorced, separated and widowed merged \\
        \texttt{pregnant} & removed & \\
        \texttt{children} & binned & $\geq 3$ merged \\
        \texttt{housing} & coarsened & shared housing categories merged \\
        \texttt{household\_\_\_5} & removed & \\
        \texttt{household\_\_\_6} & removed & \\
        \texttt{currentposition} & \multirow{2}{*}{coarsened} & registered nurse, certified nursing assistant, other with patient \\
        \texttt{position\_other} & & interaction, other without patient interaction \\
        \texttt{certifications} & removed & \\
        \texttt{nurseyears} & binned & 5 years increment, and $\geq15$ years \\
        \texttt{hours} & binned & $\leq 37.5$h, $\geq 40$h \\
        \texttt{overtime} & binned & 0h, $[1\text{h},10\text{h})$, $[10\text{h},20\text{h})$, $[20\text{h},40\text{h})$, $\geq 40$h \\
        \texttt{commute\_type} & removed & \\
        \texttt{commute\_time} & coarsened & $<15'$ and $[15',30')$ merged, $>60'$ merged \\
        \texttt{extrahours} & removed & \\
        \texttt{student} & coarsened & BSN, other (coarsened category), or \emph{no} \\
        \bottomrule
    \end{tabularx}
    }
    \caption{\revision{Demographics anonymization. We have not included in this table the variables whose values did not require changes.}
    \label{tab:data_preprocessing:surveys:demo_anonymization}}
    }
\end{table}

Data for the \revision{enrollment session baseline survey (part I), take-home baseline survey (part II)} and study-completion survey (part II) were stored in single files each. Variables were renamed to correspond to what each question measured.  After the above steps were taken, total scores for each psychological measurement were calculated (\texttt{scored} folder in \autoref{tab:main_data_record}).

\subsubsection*{Fitbit data}
Fitbit data retrieved using the Fitbit API contained separate time series for measured heart rate and step count, in addition to a daily summary of physical activity and sleep.  The heart rate data is reported on non-uniform intervals anywhere between approximately \SI{5}{\second} and \SI{15}{\minute} depending on the participants' physical activity. Occasionally, long strings of repeated identical heart rate values (usually \SI{70}{beats\per\minute}) were reported in the raw data, spanning durations typically less than 15 minutes but sometimes up to 20 hours.  Because of consumer observations that Fitbit technology sometimes incorrectly reports exactly 70bpm (see \url{https://community.fitbit.com/t5/Blaze/Blaze-s-Heart-Rate-Stuck-on-70-bpm/td-p/2727738}) and also because repeated measures of the same heart rate over several minutes is highly unlikely, these long strings were interpreted as artifacts.  Thus, sequences of at least 50 repeated identical heart rate values were replaced with \texttt{NaN} (Not a Number, equivalent to missing values). \revision{As a result, an average of $0.8\%\pm1.7\%$ of each participant's total number of heart rate samples collected were removed.} The step count, daily summary, and sleep data did not contain these long string artifacts and, therefore, were not pre-processed.

\subsubsection*{OMsignal data}
The data obtained from the OMsignal's API contained no obvious visible artifacts, and so they were not modified during the pre-processing stage.

\subsubsection*{Owl-in-One data}
The Owl-in-One devices captured packets from all Bluetooth devices broadcasting Bluetooth advertisements at Keck Hospital. We filtered all of these packets and stored only the packets coming from Minew sensors, Jelly phones, and Owl-in-Ones by filtering keywords expected to be found in the packets ("minew", "reelyActive\_RA-R436", "jelly"). These were originally stored in JSONL format, and later translated to CSV files containing only the relevant information for easier processing (details below).

\paragraph*{RSSI}\label{sec:rssi}
The RSSI information was pre-processed separately for Minew sensors, Jelly phones, and Owl-in-Ones themselves, and stored in CSV files. All MAC addresses were translated into hospital rooms or locations and formatted into a directory name as follows: \texttt{[building name]:[floor\#]:[wing/area]:[room type][room \#]}. These files also include relevant IDs (such as the participant ID associated to a Jelly phone), when appropriate. We have hashed the actual directory names to prevent making the hospital's floor plans publicly available, such that the floor number, unit, and room numbers are kept private. An example is \texttt{c25c:lounge:2fec}.

\paragraph*{Environmental data}
Bluetooth packets sent by Minew sensors contained the measured temperature and humidity, light level, or motion information in their payload.  Each packet was received by Owl-in-One devices, time stamped, and sent to the reelyActive's cloud servers where they were processed and sent to the research server. In the research server, the packets were filtered so that only packets containing Minew data were kept as environmental data.  All environmental data was further filtered so the only packets \revision{recorded} contained identifier values that also appeared on the research team's list of identifiers for all installed sensors.  \revision{Less than $0.1\%$ of the received packets contained corrupted data in the form of invalid source sensor identifiers, which is consistent with the low-energy Bluetooth (BLE) bit error rate.}  None of the other packet values were observed to be corrupted, including the measured environmental data, so no additional preprocessing was performed.

\subsubsection*{Audio}\label{sec:audio_config}
Each file contains raw audio features extracted as a combination of the Interspeech 2013 ComParE Vocalization Challenge feature set \cite{schuller2013interspeech} and openSMILE's emobase feature set \cite{eyben2013recent}. The OpenSMILE toolkit was applied in this configuration to extract acoustic low-level descriptors (LLDs) of 127 dimensions every \SI{10}{\milli\second} using either \SI{25}{\milli\second} or \SI{60}{\milli\second} frame sizes. The configuration file used to extract features is provided with the app itself (the OpenSMILE configuration file is also available at \url{https://git.io/JeiC7}).  The feature set contains prosodic measures (pitch, intensity), cepstral information (MFCCs 1-12), RASTA PLP features, spectral features (band energy between \SIrange{250}{650}{\hertz}, centroid of frequency distribution, spectral rolloffs), and other acoustic characteristics (e.g. LPC 0-7, zero crossing rate).

We did not perform any preprocessing on the raw audio before feature extraction.
To extract foreground speech information, we trained a machine learning model to learn to differentiate foreground versus background on a separate corpus collected in-house, with the same audio feature extraction hardware and software, but also with the ground truth audio, and applied it to processing the TILES-2018 Audio Data Record's raw features \cite{nadarajan2019speaker}. The output of these models is temporal foreground predictions in the interval $[0,1]$, where  values close to $1$ predict foreground. These temporal foreground predictions are also included in the TILES-2018 Audio Data Record, and described in the Data Records section.
To extract data with foreground speech, we recommend thresholding first at 0.5 a median-filtered version of the foreground speech predictions with a window length of $101$ samples (corresponding to a \SI{1}{\second} window).
A non-zero value corresponds then to a row with detected foreground speech.

For the current data release, we further curated the data by only including a subset of the features collected, and omitting filterbank features such as MFCCs and PLPs, as well as LPC features. We believe filterbanks should be released with some form of information obfuscation or encryption, as it contains potentially recoverable language information and poses privacy concerns. We intend to release privacy preserving embeddings on the filterbanks at a later stage. For information on features included in the release, refer to Section Audio.

\subsubsection*{Inference of days at work}
For convenience, we also provide an estimate of working days for all participants. This was obtained using the EMAs, as well as the data collected from the OMsignal garments and a combination of the Jelly phones and Owl-in-One data.

One of the base EMA questions was where the participant currently was (a value equal to 2 indicated currently at work). All of the participants' responses were saved into a table (each row represented a participant, each column a date). Equivalent tables were saved for days in which participants had recorded data through their OMsignal garments and through the Owl-in-Ones receiving pings from the Jelly phones.

All of this information was combined by performing a logical {or} operation between the tables. This means that if any of the sources of information regarded a given day as a day spent at work, that day was inferred as a day at work.

\section*{Data Records}
The TILES-2018 data \cite{tiles-2018} is split into two data records: the main data record, and the audio data record.  Each data set is described in detail in this section.

\subsection*{TILES-2018 Main Data Record}
The main data record is comprised of several different data streams: \texttt{fitbit}, \texttt{realizd}, \texttt{omsignal}, \texttt{owlinone}, and \texttt{surveys} (following the names of the folders in the record), and a \texttt{metadata} folder. Depending on the kind of data collected, each stream may have subfolders. These are described in the following subsections. A summary of the main data record is presented in \autoref{tab:main_data_record}. The total size of the record is about 100Gb (compressed), presented in \texttt{csv.gz} files. The files per participant are named using participants' hash-based IDs. All dates and times are in Pacific Time (PT), in the format \texttt{yyyy-mm-dd[Thh:mm:ss[.sss]]}

Detailed descriptions for all the data sources are included in each folder under a README file.

\subsubsection*{Participant summary}
\revision{
The participants were 212 hospital employees who volunteered to participate in the study. They enrolled in 1 of 3 waves of participation, each with different start and end dates (\autoref{tab:waves}). Most participants ($n=210$, 99.1\%) worked full time in the current sample. More than half of the participants were Registered Nurses ($n=113$, 54.3\%), and rest worked as Certified Nursing Assistants ($n=25$, 12.0\%), Monitor Technicians ($n=11$, 5.3\%), Physical Therapists ($n=6$, 2.9\%), Occupational Therapists ($n=2$, 1.0\%), Respiratory Therapists ($n=3$, 1.4\%), and other occupations not listed above ($n=48$, 23.1\%). The current data was collected from 146 females (rest males) and 172 individuals have received a degree higher than Bachelor’s degree. The age of the participants ranges from 21-65, with the median age at 35. 
}

\begin{table}[tb]
    \centering
    \centerline{
    \begin{tabularx}{0.8\paperwidth}{llllX}
        \toprule
        \textbf{Folders} & \textbf{Subfolders} & \textbf{Subsubfolders} & \textbf{Description} & \textbf{File Split} \\
        \midrule
        \multirow{5}{*}{\texttt{fitbit}}
            & \texttt{daily-summary} && Daily summary (aggregates, sleep) & per participant \\
            & \texttt{heart-rate} && Heart rate (PPG) & per participant \\
            & \texttt{sleep-data} && Sleep stages time series & per participant \\
            & \texttt{sleep-metadata} && Sleep periods metadata & per participant \\
            & \texttt{step-count} && Step count & per participant \\
        \midrule
        \multirow{2}{*}{\texttt{metadata}}
            & \texttt{days-at-work}          && Inferred work days from data & single file \\
            & \texttt{participant-info}              && IDs, work information & single file \\
        \midrule
        \multirow{3}{*}{\texttt{omsignal}}
            & \texttt{features}     && Recorded OMsignal features & per participant \\
            & \texttt{ecg} && OMsignal raw ECG snippets & per participant \\
            & \texttt{metadata} && Usage and data quality info   & per participant \\
        \midrule
        \multirow{6}{*}{\texttt{owlinone}}
            & \texttt{jelly}    && RSSIs to each Owl-in-One & per participant \\
            \cmidrule{2-5}
            & \multirow{3}{*}{\texttt{minew}}
                & \texttt{data}    & Temperature, humidity, and light & per device\\
                && \texttt{locations} & Locations within the hospital & single file \\
                && \texttt{rssi}    & RSSIs to each Owl-in-One & per device \\
            \cmidrule{2-5}
            & \multirow{2}{*}{\texttt{owls}}
                & \texttt{locations} & Locations within the hospital & single file \\
                && \texttt{rssi}     & RSSIs between Owl-in-Ones & per day \\
        \midrule
        \texttt{realizd} &&& Personal phone usage & per participant \\
        \midrule
        \multirow{6}{*}{\texttt{surveys}}
            & \multirow{3}{*}{\texttt{\texttt{raw}}}
                & \texttt{baseline}    & Itemized answers & per part \\
                && \texttt{EMA}        & Itemized answers & per group of scales \\
                && \texttt{post-study} & Itemized answers & single file \\
            \cmidrule{2-5}
            & \multirow{3}{*}{\texttt{scored}}
                & \texttt{baseline}        & Scored answers & per group of scales \\
                && \texttt{EMA}        & Scored answers & per scale \\
                && \texttt{post-study} & Scored answers & single file \\
        \bottomrule
    \end{tabularx}
    }
    \caption{
    TILES-2018 Main Data Record. There are five main folders containing information for each stream of data, plus a sixth folder containing participant metadata (all presented in alphabetical order). The details of each data stream (including measurements and features) are included in each of the subfolders of the data record as README files.
    }
    \label{tab:main_data_record}
\end{table}

\begin{table}[tb]
    \centering
    \centerline{
    \begin{tabularx}{\textwidth}{lllX}
        \toprule
        \textbf{Folders} & \textbf{Subfolders} & \textbf{Description} & \textbf{File split} \\
        \midrule
        \multirow{3}{*}{\texttt{audio}}
            & \texttt{raw-features} & Unfiltered audio features & \makecell[l]{many snippets\\per participant} \\
        \cmidrule{2-4}
            & \texttt{fg-predictions} & \makecell[l]{Predictions of\\foreground audio} & \makecell[l]{per participant\\and snippet} \\
        \bottomrule
    \end{tabularx}
    }
    \caption{
        TILES-2018 Audio Data Records. Audio features extracted using the Jelly phone \cite{feng2018tiles}, and extraction of snippets with foreground speech detected \cite{nadarajan2019speaker}.
    }
    \label{tab:audio_data_record}
\end{table}

\subsubsection*{Fitbit (\texttt{fitbit} folder)}
\paragraph*{\texttt{daily-summary} folder}
Each file has rows with a date and time and a set of daily summaries including resting heart rate, total calories burned, total number of steps, sleep report, and heart rate zone durations. The sleep reports provide information about sleep duration, sleep efficiency, the duration of 4 sleep stages (awake sleep, light sleep, deep sleep, REM sleep), as well as the timestamp of the start and end of the sleep. There are up to four sleep records per day. Moreover, calorie consumption and duration of 4 heart rate zones are available in Fitbit daily summaries.

\paragraph*{\texttt{heart-rate} folder}
Each file has rows with a timestamp and PPG heart rate values (beats per minute). The PPG heart rate samples are made available by the Fitbit Charge 2 sensors aggregated over intervals of less than \SI{1}{\minute}, but the time differences between two consecutive samples are non-uniform.

\paragraph*{\texttt{sleep-data} folder}
Each file has rows with the {\tt sleepId} it corresponds to in \texttt{sleep-metadata}, a timestamp and the sleep phase with its total duration in \si{\second}. Phase is either in {\tt classic} (one of {\tt asleep}, {\tt restless}, or {\tt awake}) or {\tt stages} (one of {\tt deep}, {\tt light}, {\tt rem}, or {\tt wake}). The timestamp determines the beginning of the sleeping phase.

\paragraph*{\texttt{sleep-metadata} folder}
Each file has rows for each period of sleep, and metadata for that sleep, including beginning and end, nap versus main sleep, type of inferred sleep phases ({\tt classic} or {\tt stages}), duration, and various metrics.

\paragraph*{\texttt{step-count} folder}
Each file has rows with a timestamp and step count value. In contrast to heart rate values, step count data is sampled with an interval of \SI{1}{\minute}, and reports the number of steps taken within that minute.

\subsubsection*{Metadata}
\paragraph*{\texttt{days-at-work} folder}
Contains a file for all participants. The information is presented in four tables (one per stream, plus aggregated) where each participant corresponds to a column and each row is a date in the format \texttt{yyyy-mm-dd}.

\paragraph*{\texttt{participant-info} folder}
Contains a single file with hash-based participant IDs, nursing unit(s) \revision{(if available, using the same hashing as for the Owl-in-One directories)}, and kind of shift (day or night). \revision{We have also included the dropout date if it exists.}

\subsubsection*{OMsignal (\texttt{omsignal} folder)}
\paragraph*{\texttt{ecg} folder}
Each file has raw, \SI{15}{\second}-long electrocardiogram (ECG) snippets sampled at \SI{250}{\hertz} and recorded every \SI{5}{\minute}. Each file corresponds to a single participant. Each row belongs to a single recording identified by \texttt{record\_id}, and mapping to the corresponding row in the \texttt{metadata} subfolder.

\paragraph*{\texttt{features} folder}
Each file contains rows with a timestamp and a set of physiological and physical activity measurements in real-time (aggregated and saved every second), as well as high-level descriptive features (every \SI{5}{\minute}). The real-time measurements include breathing rate, breathing depth, intensity, cadence, heart rate, RR intervals (defined as the time elapsed between two successive R waves of the QRS signal on the ECG \cite{lanfranchi2010cardiovascular}), and step count. The high-level descriptive features include statistical aggregations and derived features of real-time measurements over the \SI{5}{\minute} intervals. Examples include the average and standard deviation of the breathing rates as well as posture.

\paragraph*{\texttt{metadata} folder}
Contains one file per participant with metadata information such as dates of usage, usage time in hours, and RR coverage (ratio of successive R waves detections in time over a given time interval) for a given recording. \subsubsection*{Owl-in-One (\texttt{owlinone} folder)}
Owl-in-One data contains information from three different sources: Jelly phones (RSSIs from the Jelly phones of participants), other Owl-in-Ones (RSSIs), and Minew sensors (RSSIs and ambient information).

\paragraph*{\texttt{jelly} folder}
The \texttt{jelly} subfolder is organized with files per participant. Each file contains rows with a timestamp, a participant ID, and the directories (see Section \nameref{sec:rssi} for details) of the receiving Owl-in-Ones with corresponding RSSI values.

\paragraph*{\texttt{minew} folder}
This folder contains three subfolders:

\subparagraph*{\texttt{data} folder}
Contains one file per device whose timestamped content depends on the type of sensor:
\begin{itemize}
    \item light sensor: yes/no light detection
    \item motion sensor: acceleration in X, Y, and Z coordinates in \si{\metre\per\second\squared},
    \item temperature and humidity sensor: temperature in \SI{}{\celsius} and relative humidity in \%.
\end{itemize}

\subparagraph*{\texttt{locations} folder}
Contains a file with X and Y coordinates in \si{\metre}. The origin of the system of coordinate (\textit{i.e.}, the point $(x,y)=(0,0)$) is arbitrary so that the floor maps of the hospital are not revealed, but the pairwise distances between sensors within a same unit have been kept the same.

\subparagraph*{\texttt{rssi} folder}
This folder has one file per Minew sensor. Each file contains rows with a timestamp (sorted), the hashed directory of the receiving Owl-in-One, and the corresponding RSSI value.

\paragraph*{\texttt{owls} folder}
This folder contains two subfolders:
\subparagraph*{\texttt{locations} folder}
Contains a file with X and Y coordinates in \si{\metre} from the same arbitrary origin than the minew locations.

\subparagraph*{\texttt{rssi} folder}
The Owl-in-One files are organized by Unix time days (meaning that the cutoff is that midnight UTC). These files each contain all of the signals sent by Owl-in-Ones and received by them. Sending and receiving MAC addresses have been included, together with the sender and receivers' associated directories.
 \subsubsection*{RealizD (\texttt{realizd} folder)}
Each RealizD file describes the interaction that participants have with their smartphones. These files include a column with timestamps for initial interaction times and a column with times in seconds corresponding to the duration of the interaction. \subsubsection*{Surveys}
The \texttt{surveys} folder contains two subfolders including \texttt{raw} and \texttt{scored} surveys.

\paragraph*{\texttt{raw} folder}
This folder contains a \texttt{README} file with all the questions, and \revision{two} subfolders:

\subparagraph*{\texttt{EMAs}}
Contains three files. One file has the \revision{information} for health, personality, and job surveys plus the base daily survey in each of these. Each file contains \revision{the information on when participants were sent, started, and completed the survey}. \revision{Except for context questions, we have removed the answers to specific questions to help maintain participants' privacy}, and \revision{have kept} the information on the time until first click in each page, last click, and total time spent in each survey page. A second and third files have the responses for psychological capital and psychological flexibility respectively. These includes times at which the surveys were completed, and the total survey times. \revision{Some anonymization details may be found in \autoref{tab:data_preprocessing:surveys:ema_anonymization}, for the full details please refer to the \texttt{README} file.}

\subparagraph*{\texttt{post-study}}
Contains a single file, named with all the assessed scales. Each row corresponds to a participant's answers to each question.

\paragraph*{\texttt{scored} folder}
\subparagraph*{\texttt{baseline}}
Contains two files named with the assessed scales in each part of the survey. In each file, rows correspond to participants, and columns contain the values of the scored scales. \revision{Many of these values have been binned to help protect participants' privacy. For details, we refer the reader to \autoref{tab:data_preprocessing:surveys:demo_anonymization}}.

\subparagraph*{\texttt{EMAs}}
Each file corresponds to a scored item/scale assessed throughout the study. Each row in each file corresponds to a participant's scored answers.

\subparagraph*{\texttt{post-study}}
Contains a single file, named with all the assessed scales. Each row corresponds to a participant's answers to each question.

\subsection*{TILES-2018 Audio Data Record}
The TILES-2018 Audio Data Record contains two different kinds of files (see \autoref{tab:audio_data_record}) related to the audio features obtained as per Section Audio. Consent to publish the audio data was given by 186 out of 212 participants, as detailed in \autoref{tab:audio_consent}.

\begin{table}
	\centering
	\begin{tabular}{cSS}
	\toprule
		\textbf{Wave} & {\textbf{Gave consent} $(\bf{n})$} & {\textbf{Total} $(\bf{n})$} \\
	\midrule
		             1 &  50 &  52 \\
			         2 &  94 & 116 \\
		             3 &  42 &  44 \\
	\midrule
		\textbf{Total} & 186 & 212 \\
	\bottomrule
	\end{tabular}
	\caption{Consent given by participants to share the audio data. The consent was given at the beginning of the study through the TILES app. We have only included the data from participants who allowed their data to be shared.}
	\label{tab:audio_consent}
\end{table}

\subsubsection*{Folder Structure}
\paragraph*{\texttt{raw-features} folder} This folder is organized with subfolders per participant. The name of each data snippet in the participant subfolder is the unix time (corresponding to the time at which the recording started).

\paragraph*{\texttt{fg-predictions} folder}
This folder is arranged with subfolders per participant, like the \texttt{raw-features} folder.
Each file in a participants' subfolder is a NumPy (\texttt{.npy}) file and corresponds to a file in the \texttt{raw-features} folder.
The foreground prediction file stores an array to differentiate foreground (FG) (where FG refers to audio features generated by the participant wearing the audio badge, as opposed to background noise generated by third-party. More details can be found in \cite{nadarajan2019speaker}) and background (BG) speech activity with values indicating the likelihood of foreground speech information of each row in the corresponding file in the \texttt{raw-features} folder.

\subsubsection*{Features}
\label{sec:audio_feat}
The features are computed over overlapping frames of raw audio.
Frames lengths are typically \SI{25}{\milli\second} (and \SI{60}{\milli\second} for some features) and features are updated and recorded every \SI{10}{\milli\second} (which means roughly an overlap of 60\% for \SI{25}{\milli\second} frames and an overlap of 80\% for \SI{60}{\milli\second} frames).
Finally, some features are computed over windows of several frames.

\section*{Technical Validation}
\begin{table}[t]
    \makebox[\textwidth][c]{
    \begin{tabular}{lcrcl}
    \toprule
         \textbf{Sensor} & \makecell{\textbf{Participant}\\ \textbf{opt-in}} & \makecell{\textbf{Total}\\ \textbf{hours}} & \makecell{\textbf{Compliance}\\ \textbf{rate}} & \makecell[l]{\textbf{Definition}\\ \textbf{of compliance}}  \\
    \midrule
        Fitbit     & 208/212 (98\%) & 236,725 & 73\% & \makecell[l]{Average fraction of\\days per participant\\with > 12 hours of\\data}\\
        OMsignal   & 208/212 (98\%) &  44,240 & 60\% & \makecell[l]{Average fraction of\\ work days per\\ participant with\\ > 6 hours of data}\\
        Jelly Pro  & 184/212 (87\%) &  37,065 & 62\% & \makecell[l]{Average fraction of\\ work days per\\ participant with\\ > 6 hours of data}\\
        Owl-in-Ones &             -- &  37,065 & 98\% & \makecell[l]{Uptime of the\\ sensor network\\ (244 Owl-in-Ones)}\\
        Realizd    & 172/212 (81\%) &  --     & --   & \makecell[l]{Compliance cannot\\ be estimated due to\\ sampling scheme$^\dagger$} \\
    \bottomrule
    \end{tabular}
    }
    \caption{Sensor usage and compliance rates. Compliance is computed as the presence of data exceeding half of the measurement period per day among the participants that opted in for each sensor. Parts of this table have been reproduced from \cite{booth2019sensors}. $^\dagger$Realizd data only shows the times when phone interaction occurs, thus it is not possible to differentiate between periods with no interaction and the application not working.}
    \label{tab:compliance-sensors}
\end{table}

\subsection*{\revision{Sensor Validation}}
\revision{
In this section, we give an overview of works in the literature as well as work that we have conducted with the sensor data to validate the results.
}

\revision{
\subsubsection*{OMsignal}
We have run two studies to validate the data obtained from the OMsignal garments.
In \cite{wildman2019comparison}, we studied the differences between the heart rate data between the Fitbit and OM garments in the TILES-2018 dataset, where we observed higher correlation between Fitbit and OM garments than previous studies.
In \cite{tiwari2019comparative}, we also compared the accuracy of the Fitbit's PPG-based heart rate measurements against the OMsignal garments' ECG-based heart rate measurements. We extracted several heart rate variability (HRV) features and studied correlations with stress and anxiety.
}

\revision{
Since the discontinuation of OMsignal garments, several white papers comparing their performance with medical-grade devices were removed from the public domain. These papers showed however similar ECG quality between properly-fitted OMsignal garments and medical-grade devices.
}

\revision{
\subsubsection*{Fitbit}
There are many studies validating the data from Fitbit devices. For a full list of publications, please refer to \url{https://healthsolutions.fitbit.com/research-library/}.
}

\revision{
\subsubsection*{reelyActive Owl-in-One}
The datasheet of the reelyActive Owl-in-One version RA-H443 can be found in \cite{owl-in-one-datasheet}. In particular, each Owl-in-One uses a Texas Instruments CC2541 Bluetooth chipset, which runs reelyActive firmware \emph{without} the Bluetooth Stack to capture Bluetooth packets effectively (this information was obtained through private communication with Jeffry Dungen, Co-Founder and CEO of reelyActive).  We did not run further validation studies. However, the validation of this sensor needs to be done in two stages: At a sensor level, and at a network level. Professor Kevin Berisso from The University of Memphis performed a saturation study for the Owl-in-Ones working as receivers (work currently unpublished, kindly shared with us). In this experiment, 491 Minew beacons transmitted packets at \SI{1}{\hertz} for approximately \SI{6.5}{\hour}, with only one Owl-in-One as a receiver within \SI{2}{\meter} from all the Minew beacons. There were no lost packets reported for 303/491 Minew beacons, and 188/491 Minew beacons had reported lost packets, with a median of 71.5 lost packets for the 188 beacons with lost packets. The analysis at the network level is in the following section.
}

\revision{
\subsubsection*{TILES Audio Recorder}
We presented an analysis of the audio recorder in \cite{feng2018tiles}. TAR primarily extracts the audio features using openSMILE \cite{eyben2010opensmile}. OpenSMILE is a widely used tool for extracting a wide range of features from audio signals. To test the feature distortion from the recording device, a recording setup was proposed in \cite{feng2018tiles} to allow TAR to record speech amplified through a speaker. In this feature degradation experiment, $1000$ gender-balanced utterances from the TIMIT \cite{garofolo1993darpa} database were randomly sampled and concatenated into one file. The audio files were then played through a loud-speaker. Multiple TARs set at distances $15$~cm, $20$~cm, $25$~cm, and $30$~cm from the speaker extracted the audio features simultaneously. This experiment quantified the feature distortion by measuring the root-mean-squared error (RMSE) and cosine distance between the features extracted from the source file and recorded features. The results showed that energy-related features were sensitive to recording distance, but pitch and spectral features yield consistent patterns with different recording distances. The results also showed that errors of pitch, MFCC, and LSP were reasonably low (e.g. pitch under 10Hz), which confirmed the robustness of the feature recorded by TAR.
}

\subsection*{Data Integrity}
\begin{figure}
    \centering
    \includegraphics{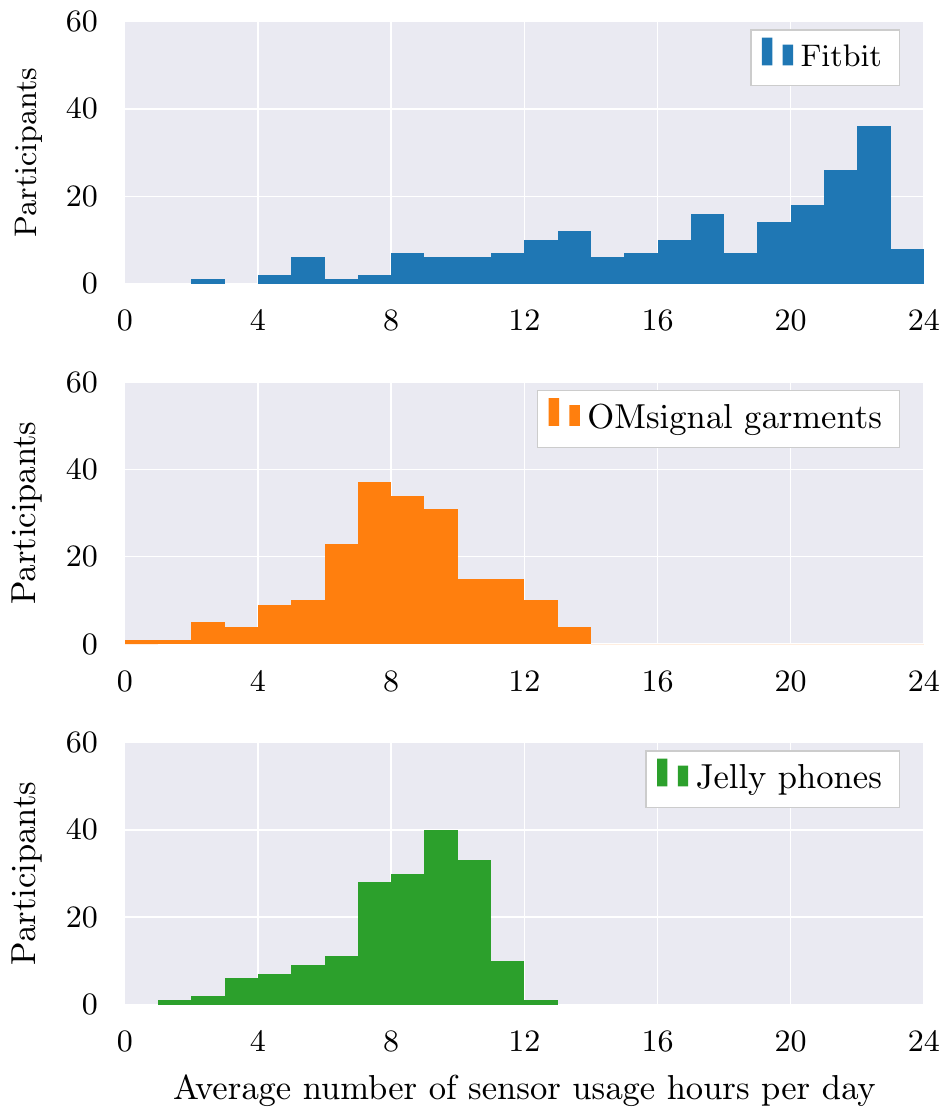}
    \caption{\revision{Daily sensor usage. These histograms show the average number of hours each participant wore the wearable sensors per day throughout the study.}
    }
    \label{fig:usage_histograms}
\tikzexternaldisable\end{figure}

The TILES-2018 data set was collected in a demanding, real-world workplace setting, where participants were asked to use wearable sensors, even though their workload and responsibilities did not change. In this scenario, the compliance rates obtained were in-line with other reported compliance rates for smaller studies, as discussed in \cite{booth2019sensors}. \autoref{tab:compliance-sensors} \revision{shows an overview of} the compliance rates for each data stream, across all participants. \revision{Opt-out reasons included privacy concerns (for the Jelly audio-feature recordings), as well as discomfort or negative skin reactions to the sensors' materials.} \autoref{fig:usage_histograms} shows histograms of the average usage hours for all the wearable sensors, across all participants. For the Jelly histogram, we have used start and stop times, which could lead to noisy estimates. \revision{\autoref{tab:sensor_compliance} shows a measure of sensor compliance in two-week intervals, where we see a slight decrease in compliance as study weeks passed.}

\revision{
\subsubsection*{Sensor Data}
}

\begin{table}[p]
    \centering
\revision{
\begin{tabular}{lcccccc}
\toprule
\textbf{Sensor} & \multicolumn{5}{c}{\textbf{Weeks}} & $\bf{p}$\textbf{-value} \\
\cmidrule{2-6}
                & \textbf{1-2} & \textbf{3-4} & \textbf{5-6} & \textbf{7-8} & \textbf{9-10} &  \\
\midrule
\textbf{Fitbit}    & 81.5\% & 79.5\% & 74.4\% & 72.2\% & 65.8\% & $<0.001$\\
\textbf{OMsignal}  & 62.5\% & 63.6\% & 64.0\% & 56.4\% & 54.2\% & $\phantom{<}0.023$\\
\textbf{Jelly Pro} & 71.3\% & 65.3\% & 68.0\% & 61.1\% & 57.9\% & $\phantom{<}0.002$\\
\bottomrule
\end{tabular}
}
\caption{\revision{Wearable sensor compliance per 2-week intervals. Compliance rate is calculated as the presence of data exceeding half of the measurement period per day among the participant that opted in for each sensor within each 2-week period. The Kruskal-Wallis test was chosen to evaluate the statistical significance of the difference between sensor compliance rate in week 1-2 and week 9-10. The presented $p$-value suggests a decreasing trend in sensor compliance rate from the beginning to the end of the study.}}
\label{tab:sensor_compliance}
\end{table}

\begin{table}[p]
\centering
\revision{
\begin{tabular}{lcS[table-format=4]S[table-format=2.2]S[table-format=4]S[table-format=2.2]S[table-format=3]S[table-format=2.2]}
\toprule
\textbf{Sensor} & \textbf{Usage} & \multicolumn{6}{c}{\textbf{\# of recordings}\quad \textbf{\%}}    \\
\cmidrule{3-8}
& & \multicolumn{2}{c}{\textbf{Wave 1}} & \multicolumn{2}{c}{\textbf{Wave 2}} & \multicolumn{2}{c}{\textbf{Wave 3}} \\
\midrule
\multirow{3}{*}{\textbf{Fitbit}} & $[8,24)$ hours & 3111 & 85.47 & 6543 & 81.99 & 2321 & 78.95  \\
                                 & $[4,\phantom{1}8)$ hours & 101 & 2.77 & 344 & 4.31 & 117 & 3.98 \\
                                 & $[0,\phantom{1}4)$ hours & 428 & 11.76 & 1093 & 13.70 & 502 & 17.07 \\
\midrule
\multirow{3}{*}{\textbf{OMsignal}} & $[4,12)$ hours & 1075 & 89.06 & 2453 & 88.33 & 765 & 89.16 \\
                                   & $[2,\phantom{1}4)$ hours & 81 & 6.71 & 180 & 6.48 & 54 & 6.29 \\
                                   & $[0,\phantom{1}2)$ hours & 51 & 4.23 & 144 & 5.19 & 39 & 4.55 \\
\bottomrule
\end{tabular}
}
\caption{\revision{Fitbit and OMsignal sensor usage. This table shows the number of recordings according to their length. Each recording corresponds to one day of data.}}
\label{tab:fitbit-om-usage}
\end{table}

\begin{table}[p]
\centering
\revision{
\begin{tabular}{ccccccc}
\toprule
\textbf{Quality} & \multicolumn{6}{c}{\textbf{\# of recordings}\quad \textbf{\%}}    \\
\cmidrule{2-7}
\textbf{Criteria} & \multicolumn{2}{c}{\textbf{Wave 1}} & \multicolumn{2}{c}{\textbf{Wave 2}} & \multicolumn{2}{c}{\textbf{Wave 3}} \\
\midrule
$[85\%, 100\%]$                     &       727 & 60.23 &       1861 & 67.01   & 513 & 59.79 \\
$[50\%, \phantom{1}85\%)$           &       285 & 23.61 &        620 & 22.33   & 200 & 23.31 \\
$[\phantom{5}0\%, \phantom{1}50\%)$ &       195 & 16.16 &        296 & 10.66   & 145 & 16.90 \\
\bottomrule
\end{tabular}
}
\caption{\revision{Data quality for OMsignal. The quality is defined as the RR coverage of the ECG signal for a given recording. Each  recording is \SI{15}{\second} long.}}
\label{tab:omsignal-integrity}
\end{table}

\begin{figure}
    \centering
    \centerline{
        \includegraphics{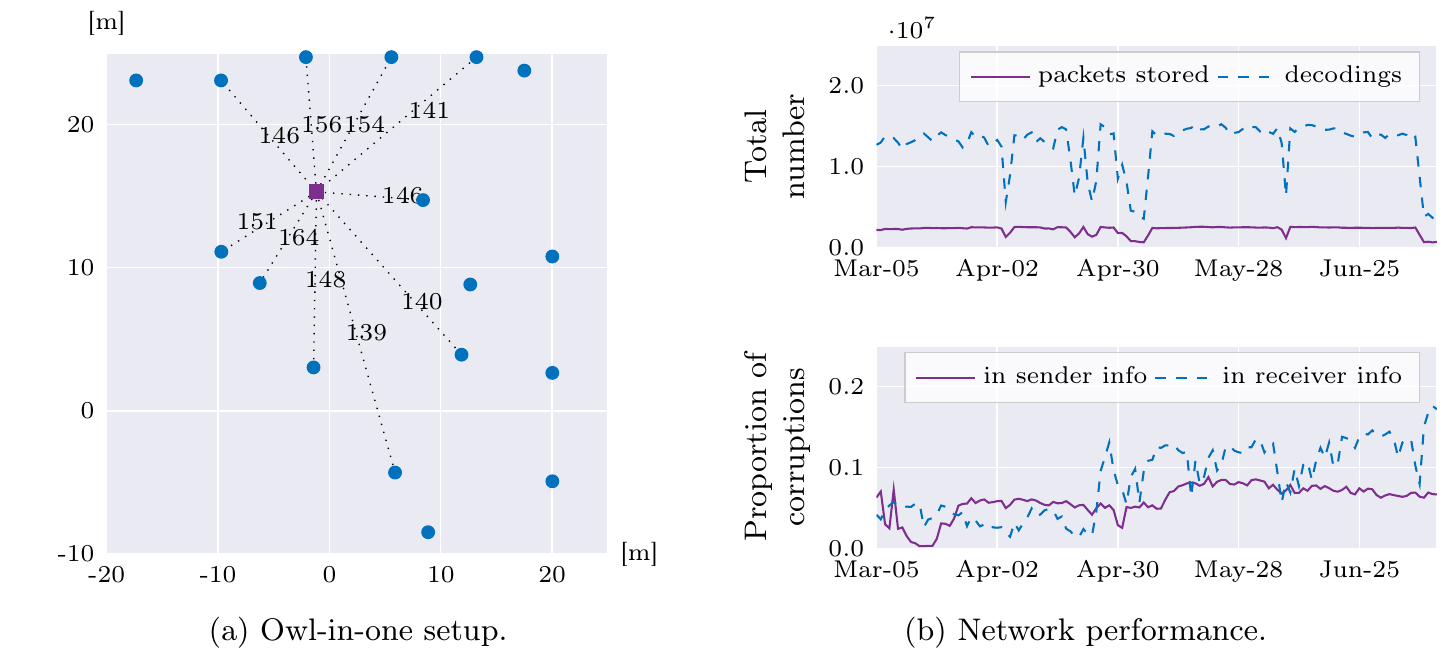}
    }
    \caption{
    \revision{
    Owl-in-One network performance. (a) Shows a typical owl setup in a nursing unit. Here, a single packet is transmitted (in this case, by purple square) and received and decoded by several other owls (as blue circles, the numbers correspond to RSSI values of decodings at the receivers). These packets contain the sender and receivers' directories. Some of these packets were processed and stored by our pipeline with corrupted information due to transmission errors. (b) (top) total number of distinct packets sent daily in the full Owl-in-One network (purple) and total number of decodings by all receiving owls (blue) and (bottom) proportion of packets whose sender directory was corrupt and therefore lost among all packets sent (purple) and receiver directory was corrupt and lost (blue) among all receiver decodings.
    }
    }
    \label{fig:owl_in_one_lost_packets}
\tikzexternaldisable\end{figure}

\revision{
\paragraph{OMsignal}{\autoref{tab:fitbit-om-usage} shows information regarding the length of the recording sessions across waves, where each recording session corresponds to the data available for a day wearing the garment. We observe that in all waves, above 88\% of the recordings are longer than \SI{4}{\hour}. \autoref{tab:omsignal-integrity}} shows the integrity of the collected data, as measured by the RR coverage of the ECG signal. In this table, we observe that about 60\% of recordings have an RR coverage of at 85\%. In both tables we observe that the usage and quality are constant across waves for the defined usability and quality. \autoref{tab:om-audio-integrity} also shows the total number of hours recorded through the OMsignal devices, and the number of participants from which these recordings were obtained.
}
\revision{
\paragraph{Fitbit}{
\autoref{tab:fitbit-om-usage} shows information regarding the daily number of hours  the Fitbit device was used, for all participants across waves. We can see a slightly downward trend through waves, with over 85\% of the recordings in Wave 1 being over 8 hours, and close to 79\% of the recordings being over 8 hours for Wave 3.
The median amount of data discarded by the Fitbit processing steps mentioned in the Data Preprocessing section was 0.3\% (approximately 1760/586681 total samples), with all but one participant having less than 7\% of their data excluded.  The last participant had approximately 20\% of their data removed during preprocessing.
}
}
\revision{
\paragraph{Owl-in-One}{\autoref{fig:owl_in_one_lost_packets} shows data integrity plots for the Owl-in-One data. \autoref{fig:owl_in_one_lost_packets} (a) shows a typical Owl-in-One layout in a nursing unit, where an Owl-in-One sends a packet that is decoded by several Owl-in-Ones in its vicinity. Here, we observe that not all owls in the vicinity are able to decode the packet, even within reach. \autoref{fig:owl_in_one_lost_packets} (b) (top) shows the total number of packets we stored in the server, as well as the total number of decodings (from the figure, 7 decodings on average per packet sent). This figure also shows some network failures during the study (dips in the discontinuous blue line). \autoref{fig:owl_in_one_lost_packets} (b) (bottom) shows the proportion of daily corrupted packets over the length of the study and the proportion of daily corrupted decodings over the length of the study. The daily average of corruptions in the sender info is 5.99\%, while the average corruptions in the receiver information is 8.49\%.}
}

\revision{
\paragraph{Audio}{
\autoref{tab:om-audio-integrity} shows the total number of hours recorded through the Jelly Pro phones, and the number of participants from which these recordings were obtained. Computing the data integrity for the audio recordings is a ver challenging problem, since the length of the recordings is variable and depends on a 2-tier sampling procedure: Uniform sampling in time over windows, and voice activity detection (VAD) over these sampling windows. If we assume that the VAD gets triggered throughout all windows for a given person in a quiet room, where only that person is speaking, the expected number of hours recorded is 120 hours for the length of the study (assuming three \SI{12}{\hour}-long units during 10 weeks).
}
}

\begin{table}
    \centering
    \makebox[\textwidth][c]{
    \begin{tabular}{llS[table-format=6.2]S[table-format=6.2]S[table-format=6.2]S[table-format=7.2]}
    \toprule
    \textbf{Sensor} &         & \textbf{Wave 1} & \textbf{Wave 2} & \textbf{Wave 3} & \textbf{Total} \\
    \midrule
    \multirow{3}{*}{\textbf{OMsignal}} & Recordings   & 2777            & 1207            & 858             & 4842    \\
                                       & Total hours  & 26421.4         & 11865.6         & 8408.9          & 46695.9 \\
                                       & Participants & 111             & 52              & 37              & 200     \\
    \midrule
    \multirow{3}{*}{\textbf{Jelly}}    & Recordings   &          410647 &          855188 &          327376 &        1593211 \\
                                       & Total hours  &         2281.37 &	      4751.04 &  	    1818.76 &  	     8851.17 \\
                                       & Participants &              51 &             101 &              42 &            194 \\
    \bottomrule

    \end{tabular}
    }
    \caption{\revision{Number of recordings and number of hours recorded for OMsignal and audio per wave. For OMsignal, each recording corresponds to a session when the participant pressed start/stop in the OM app. For audio, we show the total number of 20s-long recordings, and the total number of hours recorded.}}
    \label{tab:om-audio-integrity}
\end{table}

\revision{
\subsubsection*{Survey Data}
}
\paragraph{\revision{Baseline and Post-study}}{\autoref{tab:cronbach_alpha_baseline_post-study} shows Cronbach's $\alpha$ for the baseline and post-study surveys\revision{, as well as a validation $\alpha$ value as found in the literature}. This table shows that most of the assessments had an average $\alpha$ over 0.75, except for the agreeableness and alcohol use scales. \revision{In terms of reliability as those compared to those reported in validation studies, we have mixed results: in some cases our computed reliabilities are higher, while sometimes they are lower. However, we have increased reliabilities for all assessed scales in the post-study survey compared to the baseline survey.}}

\begin{figure}[t]
    \centering
    \centerline{
        \includegraphics{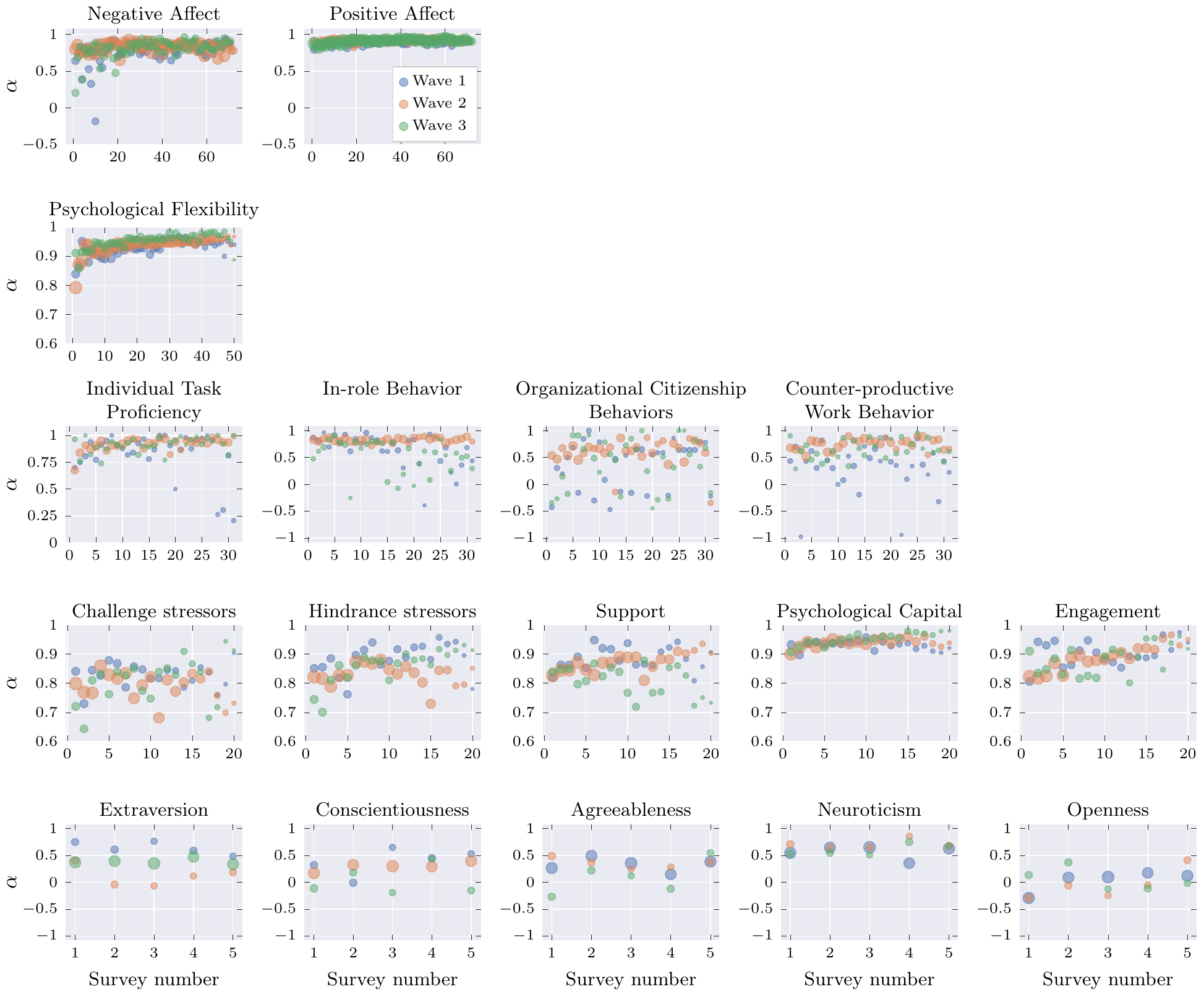}
    }
    \caption{Cronbach's $\alpha$ for each EMA. The diameter of the circles is linearly related to the number of observations for the calculation of each $\alpha$ value. Each row of scatter plots corresponds to constructs assessed the same number of times during the data collection.}
    \label{fig:alpha_emas}
\tikzexternaldisable\end{figure}

\begin{figure}
    \centering
    \includegraphics{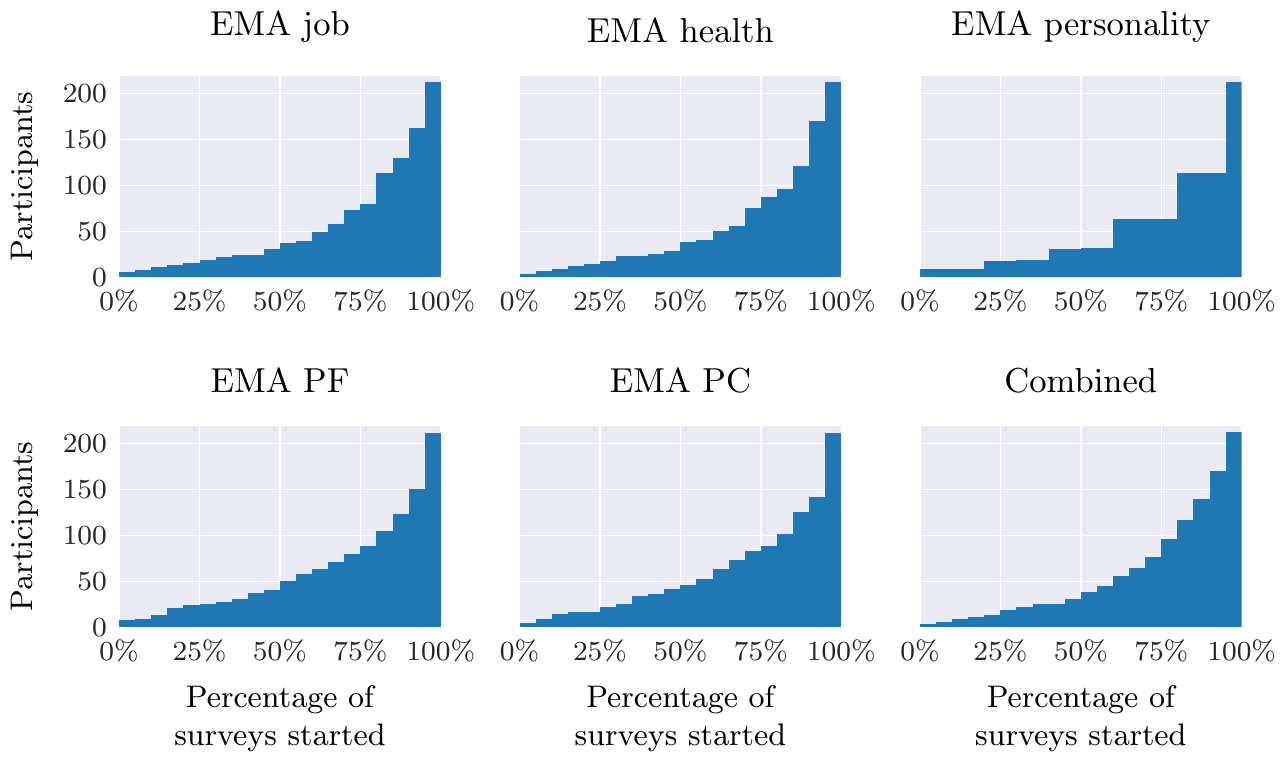}
    \caption{\revision{Number of participants with up to the specified} percentage of surveys started for each survey type. PF: Psychological flexibility; PC: Psychological capital.}
    \label{fig:surveys_started}
\tikzexternaldisable\end{figure}

\begin{figure}
    \centering
    \centerline{
        \includegraphics{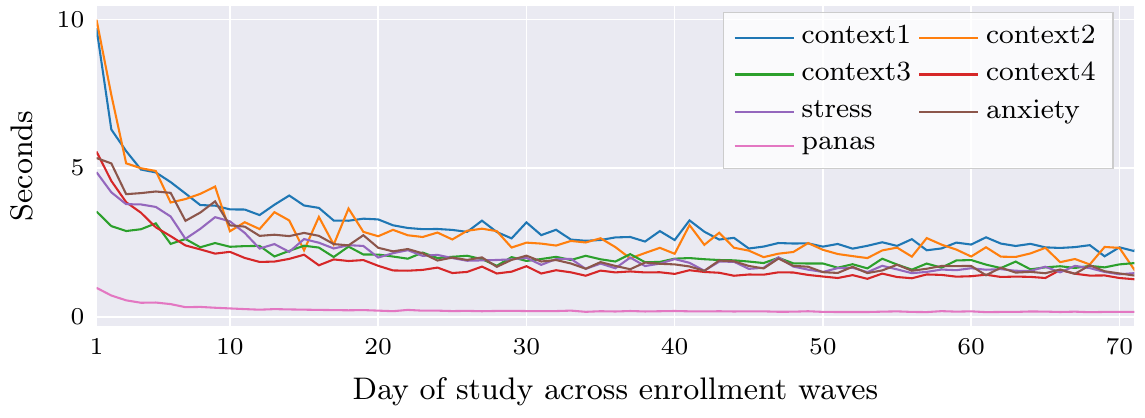}
    }
    \caption{
    \revision{
    Median EMA response times across participants. Each line shows the median of the average response times per question in a given scale for all participants. We include the baseline EMA questions present in the job, health, and personality surveys, which were asked on a daily basis.
    }
    }
    \label{fig:EMA_median_response_times_over_time}
\tikzexternaldisable\end{figure}

\paragraph{\revision{EMAs}}{\autoref{fig:alpha_emas} reports Cronbach's $\alpha$ for each construct of each EMA administered. Some of the assessed constructs show an $\alpha > 0.7$ for most of the time they were administered (challenge stressors, hindrance stressors, support, psychological capital, engagement, individual task proficiency, psychological flexibility, negative affect, and positive affect). \revision{\autoref{tab:compliance-surveys} shows the percentage of participants that opted to participate in each survey type, the average percentage of surveys per type started by participants, and the percentage of questions answered for started surveys.  The table underscores that once a participant elected to start a survey, nearly every question was answered.  \autoref{fig:surveys_started} depicts the cumulative number of participants that at least started answering at most some percentage of all surveys of each type administered throughout the study.  The histogram for each survey type illustrates that the majority of participants started responding to at least 75\% of each type of survey. \autoref{fig:EMA_median_response_times_over_time} shows the median response times for the portion of the EMAs that was asked on a daily basis. These median times correspond to the average times taken by each participant in a given scale. We observe downward trend on median time spent in questions, which we hypothesize could be associated to the participants learning the questions over time, as well as carelessly responding as questions are repeatedly asked over time. We have run unpublished analyses to detect careless responding using response times and other response-level features, however, they are not conclusive, since we observe a continuous scale of careful/careless responding among participants.}}

\subsection*{Works using the Data Set}
We have published several papers using this data, where we discuss various data processing challenges and opportunities.

In \cite{booth2019toward}, we proposed a technique for clustering and discovering patterns in proximity-based location data of hospital workers, by extracting motifs (repeating patterns) from the length of stay in each location from the proximity-based time series of locations. We used the data in this data set including locations of over 200 participants and over 240 proximity sensors during the ten weeks of the study, and discovered that rooms of similar types (e.g., patient rooms) in the hospital exhibited a unique motif signature.  The results suggest that similar motif features could be used in place of knowing the room types in advance and thus simplify very large-scale data collection.

A different approach involves using these proximity-based measurements to localize hospital workers. In our Main Data Record, we provide proximity-based information for 16 different high-traffic indoor settings. We use this information in \cite{mundnich2019bluetooth} to propose a novel indoor localization algorithm based on tools from the metric learning community.

In \cite{feng2019discovering}, we explore the usage of physiological time series collected from the Fitbit Charge 2 wristband. We particularly study how to obtain optimal-length motifs from heart rate time series to capture intuitive physiological patterns of workers in their daily lives. The results revealed that regular routine patterns, such as sleep, can be reflected through heart rate time series data.

As emphasized in \cite{booth2019sensors}, one major challenge in conducting studies in naturalistic settings relates to the quality of the data being collected with wearables. This requires the development of sensor quality metrics, missing data imputation methods, as well as quality-aware and artifact-robust parameters. To this end, we developed several such measures for breathing and heart rate time series \cite{cassani2018respiration, tiwari2019stress,tiwari2019breathing}.

The data we are publishing with the Audio Data Record proposes a new set of challenges not described in the literature before, which are related to privacy-aware audio processing in a real-world setting with sensitive information. As the Jelly phones (``audio badges'') we used recorded all foreground (egocentric audio information) and background audio of each participant's environment, in \cite{nadarajan2019speaker} we trained a machine learning model to detect foreground vs. background audio content in a different, in-house corpus which included raw audio time series alongside the same set of extracted features. We applied it to the Audio Data Record to generate foreground vs. background predictions that allow us to retrieve the egocentric information of a participant.

Finally, in \cite{kao2018discovering} we took a more global approach and used several of the data streams collected through sensors to infer self-assessments of participants, i.e., scored surveys.

\begin{table}
    \centering
    \makebox[\textwidth][c]{
    \begin{tabular}{lccc}
    \toprule
        \textbf{Variable} & \revision{\textbf{$\alpha$ validation}} & \textbf{$\alpha$ baseline} & \textbf{$\alpha$ post-study} \\
    \midrule
        Life Satisfaction (SWLS) \cite{diener1985satisfaction} & \revision{0.870} & 0.901 & 0.923 \\
    \midrule
        Perceived Stress (PSS) \cite{cohen1983global} & \revision{$\phantom{0.840 \leq } \alpha \leq 0.860$} & 0.795 & 0.830 \\
	\midrule
	Psychological Flexibility (MPFI) \cite{rolffs2018disentangling} &  &       &       \\
    	\quad Flexibility 					       & \revision{0.900} & 0.914 & 0.949 \\
    	\quad Inflexibility					       & \revision{0.902} & 0.867 & 0.875 \\
	\midrule
        Work Related Acceptance (WAAQ) \cite{bond2013work} & \revision{0.840} & 0.920 & 0.943 \\
	\midrule
        Engagement (UWES) \cite{seppala2009construct} &  & 0.913 & 0.927 \\
        \quad Vigor                  		       & \revision{$0.75 \leq \alpha \leq 0.83$} & 0.832 & 0.850 \\
        \quad Dedication            			   & \revision{$0.86 \leq \alpha \leq 0.90$} & 0.865 & 0.900 \\
        \quad Absorption                  		   & \revision{$0.82 \leq \alpha \leq 0.88$} & 0.754 & 0.765 \\
	\midrule
        Psychological Capital (PCQ) \cite{youssef2007positive} & & 0.899 & 0.934 \\
		\quad Hope             					   & \revision{$0.84 \leq \alpha \leq 0.86$} & 0.808 & 0.890 \\
        \quad Efficacy   					       & \revision{N.A.} & 0.877 & 0.907 \\
        \quad Resilience   					       & \revision{$0.770 \leq \alpha \leq 0.780$} & 0.751 & 0.787 \\
        \quad Optimism       				       & \revision{$0.780 \leq \alpha \leq 0.790$} & 0.771 & 0.871 \\
	\midrule
	Stressors \cite{rodell2009can}				   &  &       &       \\
   		\quad Challenge (CS)                       & \revision{.920} & 0.818 & 0.871 \\
   		\quad Hindrance (HS)                       & \revision{.830} & 0.781 & 0.839 \\
	\midrule
	Cognitive Ability \cite{shipleymanual}							   &  &       &      \\
		\quad Shipley Vocab (VOCAB)                & \revision{$\phantom{0.770 \leq } \alpha \leq 0.800$}  & 0.820 & N.A. \\
        \quad Shipley Abstract (ABS)               & \revision{$0.770 \leq \alpha \leq 0.910$} & 0.830 & N.A. \\
	\midrule
	Job Performance \cite{williams1991job} 							   &  &       &      \\
        \quad In-Role Behavior (IRB)               & \revision{0.910} & 0.600 & N.A. \\
        \quad Individual Task Proficiency (ITP)    & \revision{0.830} & 0.860 & N.A. \\
	\midrule
        Organizational Citizenship Behaviors (OCB) \cite{fox2012deviant} & \revision{0.890} & 0.890 & N.A. \\
	\midrule
        Counterproductive Work Behavior (IOD) \cite{bennett2000development}     &  &       &      \\
        \quad Organizational Deviance (IOD\_OD)    & \revision{0.810} & 0.820 & N.A. \\
        \quad Interpersonal Deviance (IOD\_ID)     & \revision{0.780} & 0.760 & N.A. \\
	\midrule
		Personality	(BFI-2)	\cite{soto2017next}	   &  &	   & 	  \\
        \quad Extraversion       				   & \revision{0.880} & 0.810 & N.A. \\
        \quad Agreeableness   					   & \revision{0.830} & 0.690 & N.A. \\
        \quad Conscientiousness 				   & \revision{0.860} & 0.840 & N.A. \\
        \quad Neuroticism     					   & \revision{0.900} & 0.870 & N.A. \\
        \quad Openness        				       & \revision{0.840} & 0.750 & N. A. \\
	\midrule
		Affect (PANAS-X) \cite{watson1999panas}	 					   &  &	   & 	  \\
        \quad Negative              		   & \revision{$0.700 \leq \alpha \leq 0.940$} & 0.850 & N.A. \\
        \quad Positive				       & \revision{$0.700 \leq \alpha \leq 0.940$} & 0.880 & N.A. \\
	\midrule
        Trait Anxiety (STAI) \cite{spielberger1983state} & \revision{$0.860 \leq \alpha \leq 0.950$} & 0.920 & N.A. \\
	\midrule
        Alcohol Use (AUDIT) \cite{shields2003reliability} & \revision{0.810} & 0.650 & N.A. \\
	\midrule
        Physical Activity (IPAQ)                   & \revision{0.94} & 0.870 & N.A. \\
    \midrule
       \revision{ Health (RAND)}                   & \revision{0.92} & \revision{N.A.} & \revision{N.A.} \\
    \bottomrule
    \end{tabular}
    }
    \caption{
    Cronbach's $\alpha$ for scales assessed in the baseline and post-study surveys. $\alpha$ was not calculated for GATS or the PSQI as internal consistency is not necessary for reliability on these measures. N.A.: Not assessed.
    }
    \label{tab:cronbach_alpha_baseline_post-study}
\end{table}

\begin{table}[]
    \makebox[\textwidth][c]{
    \begin{tabular}{llccl}
    \toprule
         \textbf{Survey} & \textbf{Kind} & \makecell{\textbf{Opt-in}\\($\bf{n}$ participants)} & \makecell{\revision{\textbf{Surveys started}}\\($\bf{n}$ surveys)} & \makecell{\revision{\textbf{Compliance over}}\\\revision{\textbf{Started Surveys}}\\($\bf{n}$ questions)} \\
    \midrule
        Baseline &                    &  100\% (212) & \revision{100\% (212)} & \revision{97.63\% (12296)} \\
    \midrule
                 & Job                &  100\% (212) & \revision{76.84\% (6473)} & \revision{98.89\% (147114)}  \\
                 & Health             &  97\% (204)  & \revision{77.51\% (7278)} & \revision{99.11\% (122726)}  \\
        EMAs     & Personality        &  94\% (200)  & \revision{78.64\% (1044)} & \revision{98.98\% (21346)}   \\
                 & Psych. Flexibility &  99\% (211)  & \revision{73.79\% (10390)}& \revision{98.65\% (115005)}  \\
                 & Psych. Capital     &  99\% (211)  & \revision{73.30\% (4154)} & \revision{98.70\% (88305)}   \\
    \bottomrule
    \end{tabular}
    }
    \caption{Survey participation and compliance rates \revision{over started surveys}. Compliance is measured as the percentage of answered questions in \revision{all surveys that were started}. We also include the number of started surveys.}
    \label{tab:compliance-surveys}
\end{table}

\section*{Usage Notes}
\subsection*{Data Access}
Due to privacy concerns, we request a signed \emph{Data Usage Agreement} (DUA) to grant access to all data records.
A user signing this DUA agrees to the following: (1) not de-anonymizing the data, (2) not trying to identify language content, and (3) not sharing the data record with anyone not having signed a DUA.
The document and the form to submit the signed document are available online here: \url{http://tiles-data.isi.edu}.
Once validated, the user will receive an email with the information on how to download each data record.

\subsubsection*{Main Record}
The main data record has a total size of about 100Gb.
To be mindful of the use of resources, we ask users to download this record only once.

\subsubsection*{Audio Record}
Due to the size of this data record (about \SI{305}{\giga\byte}), we provide 2 subsets of it for convenience.
A first data record is about \SI{100}{\giga\byte}, and contains only data when foreground speech has been detected.
We believe most users will be interested in this record.
The second data record of about \SI{10}{\giga\byte} is from a single user and contains all features extracted from the raw audio, unfiltered, i.e., including segment when no foreground speech has been detected.
The complete data record will only be accessible upon request, after testing has been performed on the second subset described above, to be mindful of bandwidth usage.
Same as the main records, we ask users to download each data record only once.

\subsection*{Reading the files}
We are sharing all the files as compressed comma separated values (CSV) files (\texttt{.csv.gz}), except for the foreground predictions which are stored as NumPy (\texttt{.npy}) files.
We recommend directly reading the compressed files.
This can be easily done in Python and R (examples follow in \autoref{box:loading_files}).
Note that \texttt{.csv.gz} files can also be opened directly in \texttt{LibreOffice Calc} (free software alternative to Microsoft Excel: \url{https://libreoffice.org}) without decompression.
If you \textit{need} to decompress the files, we recommend using the command line utility \texttt{gzip} or its parallelized version \texttt{pigz} for speed.

\begin{lstfloat}

\begin{lstlisting}[language=Python]
# In Python, using Pandas
import pandas as pd
df = pd.read_csv("file.csv.gz")
\end{lstlisting}

\begin{lstlisting}[language=R]
# In R, using data.table
library(data.table)
df = fread("file.csv.gz")

# or using tidyverse
library(tidyverse)
read_csv("file.csv.gz") -> df
\end{lstlisting}

\caption{Read data files in Python and R.}
\label{box:loading_files}
\end{lstfloat}

\subsection*{Data Records: Use Cases}
This data set was initially developed to model and predict self-report mental states from wearable sensors. However, we are devising more uses for it, and hope that researchers will find other uses for various aspects of the data.

\paragraph*{Multimodal Signal Processing}
This data set proposes several problems in core (multimodal) signal processing. There are several opportunities for data quality enrichment, including the denoising of the ECG snippets and Fitbit heart data, denoising of proximity information for localization, time alignment and synchronization of events from multimodal streams, and voice activity detection from breathing information. There are also new opportunities from a signal processing standpoint in the processing of longitudinal survey information.

\paragraph*{Statistical Modeling and Machine Intelligence}
This data set presents many opportunities for machine learning. The data set was initially designed to predict self-assessments of participants from sensor data. However, there are opportunities to explore the behavioral dynamics of participants throughout time, including through unsupervised learning, behavioral time series forecasting, and causal inference. Other opportunities involve spatio-temporal modeling of behavior, individualized and group-level behavioral modeling, and multitask learning of behavior patterns.

\paragraph*{Privacy}
We devise several uses of this data set for privacy researchers. Given the total number of hours of physiologic and behavioral data, an obvious use case is exploring the fingerprinting of individuals from physiologic data and behavioral patterns. We are, however, \textit{strongly} against using the data set for re-identification of specific individuals, hence a data usage agreement specifically forbidding this usage. This data set also poses new challenges and opportunities to explore venues in privacy-aware machine learning.

\paragraph*{Behavioral Sciences}
This data set poses several opportunities for research in social sciences, and specifically at the intersection of social sciences and machine learning. Some of these opportunities include the study of longitudinal survey assessments, and employee well-being within large organizations. For example, a potential avenue is to explore how data from wearable sensors relate to measures of job performance, and how they may provide new ways to explore how health and wellness impact important work behaviors. Lastly, these data provide an opportunity to examine psychometric properties of psychological measures across a 10-week period.

\subsection*{Code availability}\label{subsec:code-availability}
All code for collecting, formatting, processing, and learning on the data is made freely available at \url{https://github.com/usc-sail/tiles-dataset-release}.
Information about the code dependencies and package requirements are available in the same Github repository.

\section*{Acknowledgements}
The research is based upon work supported by the Office of the Director of National Intelligence (ODNI), Intelligence Advanced Research Projects Activity (IARPA), via IARPA Contract No 2017-17042800005. The views and conclusions contained herein are those of the authors and should not be interpreted as necessarily representing the official policies or endorsements, either expressed or implied, of the ODNI, IARPA, or the U.S. Government. The U.S. Government is authorized to reproduce and distribute reprints for Governmental purposes notwithstanding any copyright annotation thereon.

We thank Felipe Osorno and Brooke Baldwin-Rodríguez from USC Keck Hospital for their assistance, and all the volunteers that decided to be part of the study. We thank Luca Foschini from Evidation Health for his helpful insights during the study design.
We also thank Jeffrey Dungen from reelyActive and the team from OMsignal for their amazing support.
We finally thank Cynthia Begay and multiple Research Assistants who helped with the implementation and day-to-day logistics.

\section*{Author contributions}
\noindent\textbf{Karel Mundnich}: Study design, sensor testing and selection, sensor installation and management, Owl-in-One data preprocessing. Data curation. Main author of this paper.\\
\textbf{Brandon M. Booth}: Study design, sensor testing and selection, sensor installation, data preprocessing. Contributed to Sensor Installation and Data Preprocessing sections of this paper as well as the data set website and code availability.\\
\textbf{Michelle L'Hommedieu}: Assisted with study design, survey design, and recruitment. Study implementation. Contributed to the Methods section of this paper.\\
\textbf{Tiantian Feng}: Sensors testing and selection, design and implementation of the audio recorder. Contributed to the Data Records section and Technical Validation of this paper.\\
\textbf{Benjamin Girault}: Data management, preprocessing, and curation, repository selection and management. Contributed to Methods, Data Records, Technical Validation, and Usage Notes sections of this paper.\\
\textbf{Justin L'Hommedieu}: Study implementation, survey scoring. Contributed to the Methods section of this paper.\\
\textbf{Mackenzie Wildman}: Developed data collection pipeline, study companion application (TILES app), and incentives scheme. Contributed to the Methods section of this paper. \\
\textbf{Sophia Skaaden}: Assisted with study implementation. Contributed to the Enrollment Session subsection within the Methods section.\\
\textbf{Amrutha Nadarajan}: Study design, sensor selection and management, quality checks for audio and OMsignal. Contributed to the Audio Record section.\\
\textbf{Jennifer Villatte}: Study design and framing, survey design. Contributed to the Background \& Summary, and the structure of the Main Data Record.\\
\textbf{Tiago H. Falk}: Study design, sensor selection, audio feature selection, and multimodal data analysis. Contributed to the overall manuscript.\\
\textbf{Kristina Lerman}: Study conception, design, and coordination. Data validation and analysis. Reviewed the manuscript.\\
\textbf{Emilio Ferrara}: Study conception and design, multimodal data analysis and coordinator of machine learning modeling. Co-PI of the study. Contributed to the overall manuscript.\\
\textbf{Shrikanth Narayanan}: Study conception, design and implementation, study coordination and management, and leader of the study. Contributed to the overall manuscript.\\
All authors read and approved the final manuscript.

\section*{Competing interests}
The authors declare no competing interests.

\FloatBarrier
\section*{Figures and figures legends}
\tikzexternalenable

\tikzexternalenable

\tikzexternalenable

\tikzexternalenable

\tikzexternalenable

\tikzexternalenable

\tikzexternalenable

\tikzexternalenable

\FloatBarrier

\end{document}